\documentclass[fleqn,usenatbib]{rasti}

\usepackage{newtxtext,newtxmath}

\usepackage[T1]{fontenc}

\DeclareRobustCommand{\VAN}[3]{#2}
\let\VANthebibliography\thebibliography
\def\thebibliography{\DeclareRobustCommand{\VAN}[3]{##3}\VANthebibliography}

\usepackage{graphicx}	
\usepackage{amsmath}	
\usepackage{pifont}
\usepackage{multirow}
\usepackage{marvosym}

\newcommand{\cmark}{\ding{70}}%
\newcommand{\xmark}{\Circpipe}%

\title[Evaluating RFI mitigation algorithms in single pulse search]{Evaluating the effectiveness of radio frequency interference removal algorithms for single pulse searches}

\author[Hombal et al.]{
R. S. Hombal$^{1}$\thanks{E-mail: raghuttamshreepadraj.hombal@manchester.ac.uk},
L. Levin$^{1}$, 
B. W. Stappers$^{1}$,  
M. Droog$^{3}$, 
A. Karastergiou$^{2}$,
D. Lumbaa$^{4}$,
\newauthor M. B. Mickaliger$^{1}$, 
A. Naidu$^{2}$, 
K. M. Rajwade$^{2}$,
J. Sepulveda$^{1}$ 
B. Shaw$^{1}$, 
S. Singh$^{1}$,
T. Prabu$^{5}$
\\
$^{1}$Jodrell Bank Centre for Astrophysics, Department of Physics and Astronomy, The University of Manchester, Manchester M13 9PL, UK\\
$^{2}$Department of Astrophysics, University of Oxford, Denys Wilkinson Building, Keble Road, Oxford OX1 3RH, UK\\
$^{3}$Covnetics Ltd, Eliot Park Innovation Centre, 4 Barling Way, Nuneaton CV10 7RH, UK\\
$^{4}$Akkodis UK Limited, New Filton House, Filton, Bristol BS34 7QQ, UK\\
$^{5}$Raman Research Institute, Sadashivanagar, Bengaluru 560080, India
}
\date{Accepted XXX. Received YYY; in original form ZZZ}

\pubyear{2025}

\begin{document}
\label{firstpage}
\pagerange{\pageref{firstpage}--\pageref{lastpage}}
\maketitle

\begin{abstract}
 Radio Frequency Interference (RFI), the presence of artificial and/or terrestrial signals in astronomical data, poses a great challenge to the search for pulsars and radio transients, such as Rotating Radio Transients (RRATs) and Fast Radio Bursts (FRBs), by obscuring or distorting the signal of interest and resulting in large numbers of erroneous detections.  RFI mitigation algorithms aim to remove this interference and improve the chance of detection of transients, but with the growing number of techniques, selecting the most appropriate method for a given survey can be problematic. The choice of method is particularly important in real-time searches planned for next-generation telescopes such as those of the SKAO, where there is no possibility to reprocess the data. In this paper, we explore the algorithm selection problem by injecting pulses into data which simulates several RFI environments. A set of these files is then cleaned using RFI mitigation algorithms and run through a single pulse search pipeline to analyse the recovery of the injected pulses. We examine the recovery of the injected single pulses with an emphasis on a number of cases spanning a range of pulse brightness, width and dispersion measure. The efficacy and side effects of a few popular RFI excision methods, namely IQRM, SKF, and ZDMF are evaluated.
\end{abstract}




\section{Introduction}
\label{sec:introduction}
The detection of single short-duration pulses of radio emission has led to the discovery of many astronomical objects such as pulsars and rotating radio transients \citep{pulsar_discovery,rrat_discovery}. Pulsars are highly magnetised, rapidly rotating neutron stars emitting radio waves from their magnetic poles, detected as a pulse of emission if the beam crosses the observer's line of sight. Rotating radio transients (RRATs) are a subclass of neutron stars which emit short but infrequent radio bursts compared to the general pulsar population. \citet{sps_technique} proposed a method for detecting short-duration radio pulses that is now widely used to explore various regions of the transient parameter space, spanning a range of different timescales and luminosities. The use of this technique serendipitously led to the discovery of a distinct class of phenomena known as fast radio bursts \citep[FRBs;][]{frb_discovery}. FRBs are characterised by short-duration but very bright radio bursts, whose origin remains unknown. Recent discoveries \citep[][]{CHIME_FRB_magnetar, CHIME_STARE2_FRB_Magnetar} and the identification of a population of repeaters \citep[e.g.][]{repeating_frb_chime} suggest connections between FRBs and magnetars might explain at least some of the population. However whether this is the only type of source that can produce FRBs is still unknown. Many models for the origin of FRBs can be found in \citet{Platts_FRB_cat} and more details on FRBs can be found in these reviews \citet{Petroff_2022_frb_review, zhang_phy_frb, review_lorimer_frb}.
Discoveries like FRBs motivate astronomers to explore all parts of the parameter space of possible radio transients, not only unexplored regions but also to revisit the previously studied regime with improved sensitivity and algorithms. However, with increased radio frequency interference (RFI) in the observable radio spectrum, this becomes a challenge as signals from transients can be obscured. Any artificial, naturally occurring or any non-white-noise-like signal that can negatively impact astronomical observations is often referred to as RFI. The observed signal at the end of the telescope signal chain is the sum of the contributions of the astronomical source, thermal noise contributed by the parts of the instrument itself, backgrounds such as the sky and the ground, and RFI. Typical power densities of astronomical sources are in the range of $-220\,\mathrm{dB~W~m^{-2}}$ to $-150\,\mathrm{dB~ W~ m^{-2}}$ (equivalent to $0.1\,\mathrm{mJy} - 100\,\mathrm{Jy}$ when observed with a frequency bandwidth of $100\,\mathrm{MHz}$). Radio telescopes are therefore required to be extremely sensitive, as even the strongest astronomical signal at a frequency of $300\,\mathrm{MHz}$, $10^5\,\mathrm{Jy}$, is still approximately 10$^6$ times weaker at the telescope than the interference from a typical communication transmitter \citep[see][]{rfi_level}. 

In addition to their strength, RFI can exhibit combinations of duty cycle, bandwidth variations, unusual shapes, chirp-like structures, complex modulations, and frequency-dependent variations. Natural phenomena like lightning can affect wide frequency bands, whereas transmitters such as mobile and telecommunication systems, frequency-modulated and Amateur (HAM) radio transmitters, radars, and others occupy designated frequency bands. These narrowband transmitters usually transmit modulated signals to facilitate longer propagation ranges, which may affect adjacent frequency bands to those intended for communication. In addition to this, there can also be signatures of Doppler-shifted RFI, such as that from satellites directly or reflected off aeroplanes.

With the growing number of RFI sources, it is becoming paramount to mitigate the effects of RFI. 
When searching for radio transients, the relative strength and transient, and/or modulated, nature of RFI could lead to the RFI being reported as real astrophysical sources (false positives) or result in measuring inaccurate source parameters, such as strength and/or width, or complete obscuration of a real astrophysical signal \citep[][]{perytons}. Sometimes, they might also imitate spectral lines \citep[][]{rfi_bann}.

RFI can be mitigated using several methods, such as frequency rejection and spatial filtering \citep[e.g.][]{rfi_bann}. Each method has its own advantages and limitations in its ability to excise RFI. The frequency rejection method excises RFI by applying a mask or notch filter to eliminate frequency channels that are either known a priori or predicted to be contaminated with RFI. The spatial filtering method localises the RFI emission using a reference antenna pointing off-source. It suppresses the unwanted signal by nulling the synthesised antenna pattern that coincides with incoming RFI \citep[see][]{vanArdenne_2000_spatial_filtering}. These methods are effective, but in this paper, we wanted to consider the methods that are directly applied to the time-frequency space, henceforth called dynamic spectra.

Several RFI Mitigation (RFIM) techniques have been designed to remove RFI signatures from dynamic spectra: Sum-threshold algorithm \citep{sumthreshold}, Zero-DM filter \citep{zdf}, Spectral Kurtosis filter \citep{skf2007,skf}, Inter Quartile Range mitigation \citep{iqrm}, and Zero-DM Matched filtering \citep{zdmf} are a few popular algorithms that are often used (details in section~\ref{sec:rfim_details}). Recently, deep learning algorithms have also been explored to excise RFI \citep[e.g.][]{ml_rfim_wang_2020, ml_rfim_yang_2020, ml_rfim_benjamin_2022, ml_rfim_sadr_2020}.

Generally, the appropriate RFIM algorithms and parameters are chosen based on the performance they achieve when applied to observational data, but to the best of our knowledge, a comparison using a set of controlled parameters and input data has not been explored in the literature. The enormous data rates and the need for rapid follow-up of fast transients, means that telescopes have to run data processing pipelines in real-time, e.g. for the SKAO and its precursors \citep[see e.g.][]{Macquart_2010_craft_askap_pipeline, Sanidas_2017_meertrap_paper,ska_pss}, therefore, care must be taken when choosing the most appropriate algorithm as the raw data is no longer available. In addition to this, RFIM algorithms not only have to remove RFI but also have to preserve the intrinsic properties of the single pulse events as much as possible. In this work, we present a method for optimising the set of RFIM algorithms and their parameters to select an appropriate combination of RFIM algorithms in order to minimise missing candidates and the number of false positives.

\citet{rfi_compare_shapiro_wilks} evaluates the effectiveness of algorithms capable of cleaning channelised voltage data (using Median Absolute Deviation) and power spectral density (Spectral Kurtosis), by comparing the resulting signal-to-noise (S/N) of the folded pulse profile of a test pulsar. They compared the algorithms that work on pre-detection data (stage where the data are raw voltages), whereas we are working with post-detection data (a stage where the data are in units of power), a constraint common to many transient searches using total power or Stokes data, where the ability to access and clean raw voltages is lost. 
\citet{iqrm} compared their proposed algorithm to other existing ones using post-detection observational data from telescopes. Using observational data is useful for comparing algorithms, but it is more difficult to determine whether one has recovered the expected signals and parameters of the astrophysical sources that might be included. One could inject pulses into real data, however, there may be uncertainty because the data may include underlying RFI, baseline variations and perhaps other instrumental effects that we are not in control of.

We proceed with the assumption that if all the unknown underlying RFI instances mentioned above were absent, the data would be noise-like. In our approach, we therefore conduct tests on dynamic spectra whose contents are completely under our control with minimal uncertainties. This provides a way to directly measure the effectiveness of the algorithms at detecting the known input pulses. This work was in part motivated by that of \citet{analysis_van}, who assessed the effect of non-stationary Gaussian noise and RFI on standard pulsar search pipelines and their ability to detect pulsars.

\section{Methodology}
\label{sec:process}
Our approach to evaluating the effectiveness of RFI mitigation algorithms comprises three stages: generating test vectors, applying the RFIM algorithm to remove the injected RFI, and performing a search to recover the pulses injected. A test vector is a controlled representation of the data that would be presented to a search pipeline and can be used to evaluate the response of the search pipeline. We note that we use detection fraction as our metric here, as we are interested in seeing whether the pulses are recovered during the real-time search rather than investigating the accuracy of the pulse parameters detected. Typically, real-time search pipelines for fast transients will preserve a small amount of complex voltage data at the time of the detected pulse, and these data can be used at a later stage to get the best possible parameters for the detected pulses. We also run the searches over a range of dispersion measures (DMs) that are representative of a typical single pulse search campaign. This is because we want to test whether the presence, or imperfect removal of RFI can affect the detectability of the injected pulses and/or result in them being detected at the incorrect DM, width, time of arrival, and S/N. We note that the imperfect removal of RFI might also lead to a large number of false positives, which results in the search pipelines missing the astrophysical pulses, or becoming non-real-time (see Appendix \ref{sec:errors}).

\begin{table}
	\centering
	\caption{Single pulse parameters used to create test vector filterbanks.}
	\label{tab:no_rfi_testvectors}
	\begin{tabular}{|l|l|} 
		\hline
		Parameter & Values\\
		\hline
        Integration time & 60 s \\
        Sampling time & $\mathrm{64~\mu s}$ \\
        Number of frequency channels & 4096 \\
        Frequency of highest channel & 1670 MHz \\
        Channel bandwidth & 78.125 kHz\\
        \hline
		\multirow{8}{*}{DM ($\mathrm{pc/cm^3}$)} & 10 \\
                                                & 20 \\
                                                & 100 \\
                                                & 150 \\
                                                & 300 \\
                                                & 500 \\
                                                & 1000 \\
                                                & 3000 \\ 
        \hline
		\multirow{4}{*}{Pulse widths ($\mathrm{ms}$)} & 8\\
                                                    & 40\\
                                                    & 80\\
                                                    & 800\\
        \hline
		\multirow{5}{*}{Signal-to-noise} & 9.1\\
                                            & 14.1\\
                                            & 42.4\\
                                            & 84.9\\
                                            & 141.4\\
		\hline
	\end{tabular}
\end{table}

\subsection{Test vectors}

\begin{figure*}
    \centering
    \includegraphics[width=0.95\linewidth]{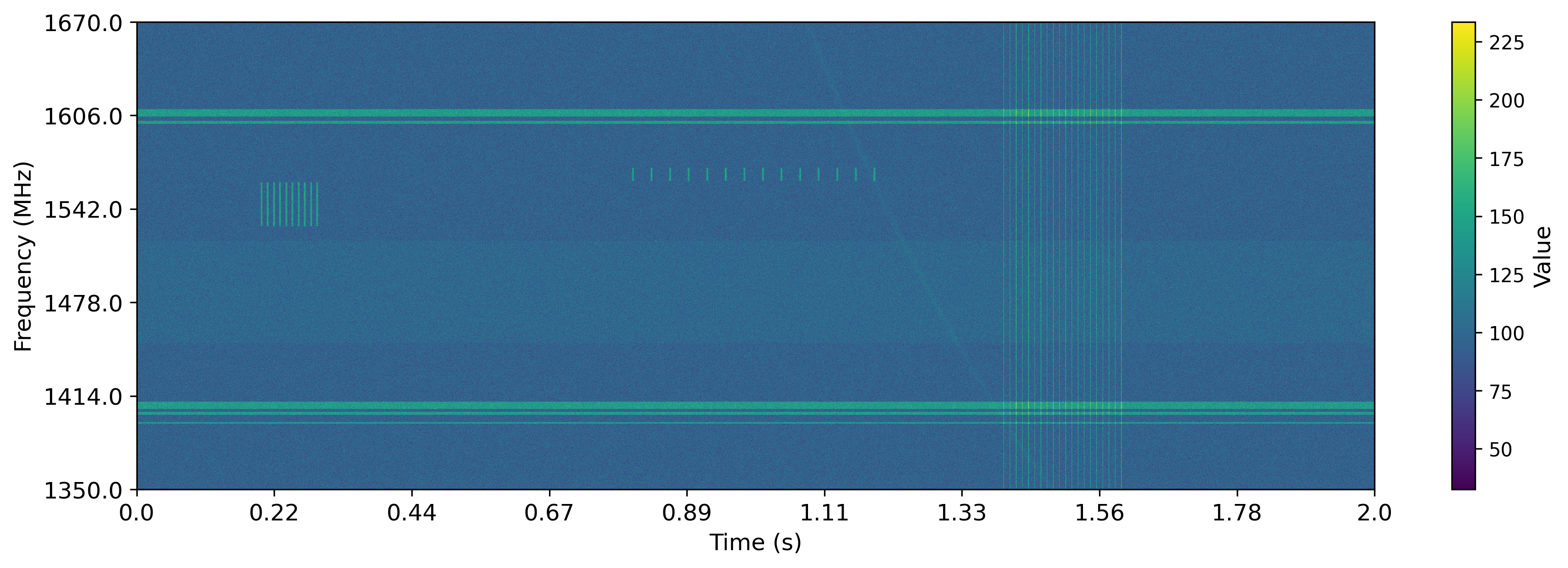}
    \caption{Data in a filterbank file that contains a simulated radio transient with a DM of $500\,\mathrm{pc/cm^2}$ with a S/N of 141 and added RFI. This test vector shows strong RFI ($\mathrm{2\sigma}$) of all the types, i.e. narrowband constant, narrowband periodic and broadband periodic RFI. A Gaussian-shaped, dispersed pulse can be seen sweeping from around $1.11\,\mathrm{s}$ in time in the highest frequency channel ($1670\,\mathrm{MHz}$) up to $1.5\,\mathrm{s}$ in the lowest frequency channel ($1350\,\mathrm{MHz}$). The structures around $\mathrm{1606~MHz}$ and $\mathrm{1410~MHz}$ mimic the constant narrowband RFI that can be seen from fixed-frequency terrestrial transmitters, which affect the entire observation. The periodic structures affecting the full bandwidth around $1.5 ~\mathrm{s}$ in time are similar to those of lightning or some other periodic broadband signal. The periodic structures affecting a narrowband of frequencies from $\mathrm{0.22~s}$ and $\mathrm{0.7~s}$ are like those produced by a communication satellite or mobile phones. The data has been downsampled by a factor of 2 in time and frequency to make the pulse stand out.} 
    \label{fig:realistic_rfi_filterbank}
\end{figure*}

The process of generating test vectors is summarised in the following steps:
\begin{itemize}
    \item Generating a noise file,
    \item Injecting pulses of various DMs, widths and S/N as detailed in Table \ref{tab:no_rfi_testvectors}, which results in 160 test vectors,
    \item Injecting 21 realisations of RFI into each of these test vectors, leading to a total of 3360 test vectors.
\end{itemize}
This three-step process is explained in detail in the sections below.

\subsubsection*{Generating a noise file}
Table \ref{tab:no_rfi_testvectors} contains the telescope parameters and the properties of the pulse used to create these test vectors. The telescope parameters were chosen to be similar to those expected for band 2 of SKA-Mid. For the experiment described in this paper, as discussed above, we assume that the noise in the generated data, prior to insertion of any RFI or astrophysical pulses, is described by a Gaussian distribution\footnote{While we model the noise as Gaussian for simplicity, the actual underlying distribution could be different, such as $\mathrm{\chi^2}$. This choice does not impact the study's conclusions, as the RFI mitigation algorithms employed for our tests are agnostic to the form of the noise distribution.}. The test vectors are therefore generated using 8-bit unsigned integers drawn from normally distributed noise with fixed mean and standard deviation. These noise files are saved in the \texttt{SIGPROC} filterbank format, which is used as a standard data format for transient search pipelines globally \citep[see][]{sigproc}.

\subsubsection*{Pulse Injection}
A pulsar of desired S/N, with a DM and a Gaussian-shaped pulse of chosen width, is injected into the filterbank noise file using a package called \texttt{filtools} \citep{filtool_cite} for each combination of these pulse parameters. 
The pulsar is injected periodically with a fixed S/N, determined by the requested S/N of the integrated pulsar signal. The way \texttt{filtools} works means that it injects pulses that are identical to each other. Thus, they will differ only due to the contribution from the underlying noise. The pulses are injected with a fixed period of $\mathrm{8~s}$ and without intra-channel DM smearing. The period is chosen so that each test vector contains 6 individual pulses. 
We have injected pulses with a constant amplitude across all frequencies, and without any temporal variation, because we wanted to perform a first-order experiment, and including more free parameters such as spectral features, scintillation and temporal variation would mean additional complexity. We can use the known arrival time of the pulse when determining whether the search process has correctly identified the input pulse.

The ranges of parameters are chosen in such a way that they cover as many cases as possible for a real astrophysical pulse that might be detected in the future. The range is sampled on a log scale to capture as wide a set of parameters as we can in a definite and manageable number of possible values. Three sets of test vectors were generated that contained different realisations of RFI. The first set of test vectors contains white noise with pulses and no RFI is injected (referred to as TVS-1 henceforth). The test vectors of sets 2 and 3 were generated by adding RFI along with the pulse parameters as shown in Table~\ref{tab:no_rfi_testvectors}.

\subsubsection*{RFI Injection}
The RFI instances we inject into the filterbank are designed to mimic real-life RFI sources in a simplified but representative way, which are sufficient to test how well an algorithm will excise corresponding real-world RFI. We chose to do this to have control over the nature of RFI in the data. Capturing the complete range of possible RFI manifestations in a finite set of simulated observations is a challenging task. Any instance of RFI in our test vectors can be defined by an amplitude and the number of frequency channels and time samples it spans. Additionally, we use a period and duty cycle for some of these RFI instances to introduce periodicity and make the injection process easier. As most RFI does not have sharp edges in time and/or frequency that would match our time and frequency sampling, the edges of the RFI instances are smoothed by approximately $\mathrm{0.8~MHz}$ in frequency and $\mathrm{640~\mu s}$ in time. This is done by convolving the RFI instances with a mask defined in a two-dimensional array \citep{scipy-virtanen}.

As mentioned in section \ref{sec:introduction}, RFI can manifest in different forms from various sources. Despite the complexity, RFI can generally be categorised into four types: constant broadband, constant narrowband, periodic broadband and periodic narrowband. We note that we consider periodic broadband RFI to capture the nature of constant broadband RFI as well, because the algorithm that we use for broadband RFI mitigation is insensitive to the temporal width of the RFI.  All types of RFI we inject are represented in Fig. \ref{fig:realistic_rfi_filterbank}. It shows an example of a piece of a filterbank file as a dynamic spectrum (time against frequency with colour scale corresponding to the strength of the signal) with simulated RFI and an injected pulse. 

The second set of test vectors (referred to as TVS-2 henceforth) contains a combination of these narrowband, broadband and periodic RFI. A combination can be made with each type with varied amplitudes, thus creating a combination of 8 test vectors for a given DM, S/N and pulse width as shown in Table~\ref{tab:realistic_rfi_testvectors}. When an instance of RFI is injected, it has a uniform strength which corresponds to an increase in the mean by $\mathrm{n \sigma}$ per frequency channel per time sample, where $\mathrm{n}$ can be different for different types of RFI in a given test vector. A total of $5\%$ of the frequency channels and time samples in TVS-2 are contaminated by RFI.

\begin{table}
	\centering
	\caption{RFI environments simulated for the TVS-2 test vectors. The symbols show the strength of RFI, where \cmark~ represents $\mathrm{2\sigma}$ (Strong RFI) and \xmark~ represents $\mathrm{0.5\sigma}$ (Weak RFI).}
	\label{tab:realistic_rfi_testvectors}
	\begin{tabular}{|l|c|c|c|} 
		\hline
		Type & Narrowband & Broadband & Periodic\\
		\hline
		1 & \xmark & \xmark & \xmark\\
        2 & \xmark & \xmark & \cmark\\
        3 & \xmark & \cmark & \xmark\\
        4 & \xmark & \cmark & \cmark\\
        5 & \cmark & \xmark & \xmark\\
        6 & \cmark & \xmark & \cmark\\
        7 & \cmark & \cmark & \xmark\\
        8 & \cmark & \cmark & \cmark\\
        \hline
	\end{tabular}
\end{table}

\begin{table}
	\centering
	\caption{The RFI parameters used to simulate the RFI environment used to create TVS-3 test vectors. The percentage of affected samples refers to the fraction of affected frequency channels in the case of narrowband RFI and time samples in the case of broadband RFI. TVS-3A comprises the narrowband RFI realisations, and TVS-3B the broadband RFI ones.}
	\label{tab:extreme_rfi_testvectors}
	\begin{tabular}{|l|c|c|} 
		\hline
		Type of RFI & \% of affected samples & Strength\\
		\hline
		Narrowband & 25\% & $\mathrm{2\sigma}$\\
        Narrowband & 25\% & $\mathrm{1\sigma}$\\
        Narrowband & 25\% & $\mathrm{0.5\sigma}$\\
        Narrowband & 10\% & $\mathrm{2\sigma}$\\
        Narrowband & 10\% & $\mathrm{1\sigma}$\\
        Narrowband & 10\% & $\mathrm{0.5\sigma}$\\
        Broadband & 25\% & $\mathrm{2\sigma}$\\
        Broadband & 25\% & $\mathrm{1\sigma}$\\
        Broadband & 25\% & $\mathrm{0.5\sigma}$\\
        Broadband & 10\% & $\mathrm{2\sigma}$\\
        Broadband & 10\% & $\mathrm{1\sigma}$\\
        Broadband & 10\% & $\mathrm{0.5\sigma}$\\
        \hline
	\end{tabular}
\end{table}

The third set of test vectors has two subsets, referred to as TVS-3A and TVS-3B henceforth. TVS-3A contains only evenly spaced narrowband periodic RFI, and TVS-3B contains only broadband RFI, contaminating all frequency channels of a few time samples randomly repeated over time. The maximum number of affected frequency channels and time samples was restricted to $25\%$ of the total number of frequency channels or time samples. Note that each of the test vectors in a subset contains the same kind of RFI, for instance, in TVS-3B, every test vector contains RFI affecting the same group of frequency channels (see fig.~\ref{fig:testvector_filterbank_broadband} and \ref{fig:testvector_filterbank_narrowband}). Table~\ref{tab:extreme_rfi_testvectors} shows all the combinations of values of RFI parameters used to generate test vector sets TVS-3A and TVS-3B containing RFI.

Table~\ref{tab:tvs-truth-table} summarises all the test vector sets and their contents. A Python-based script, \texttt{Generator.py}\footnote{https://gitlab.com/ska-telescope/pss/ska-pss-test-vector-generator/ . The Gaussian noise generator uses \texttt{123123} as the seed value to generate white noise to inject pulses and RFI into.} was used to automate the tasks mentioned in the section above.

\begin{table}
    \centering
    \caption{Truth table of contents in each set of test vectors}
    \begin{tabular}{|p{1cm}|p{1.2cm}|p{1.2cm}|p{1.2cm}|p{1.2cm}|}
    \hline
         & Pulses & Periodic broadband & Periodic narrowband & Persistent narrowband \\
         \hline
        TVS-1 & \ding{51} & \ding{53} & \ding{53} & \ding{53} \\
        TVS-2 & \ding{51} & \ding{51} & \ding{51} & \ding{51} \\
        TVS-3A & \ding{51} & \ding{51} & \ding{53} & \ding{53} \\
        TVS-3B & \ding{51} & \ding{53} & \ding{51} & \ding{51} \\
        \hline
        
    \end{tabular}
    \label{tab:tvs-truth-table}
\end{table}

\subsection{RFIM algorithms}
\label{sec:rfim_details}
\begin{figure*}
    \centering
    \includegraphics[width=0.95\linewidth]{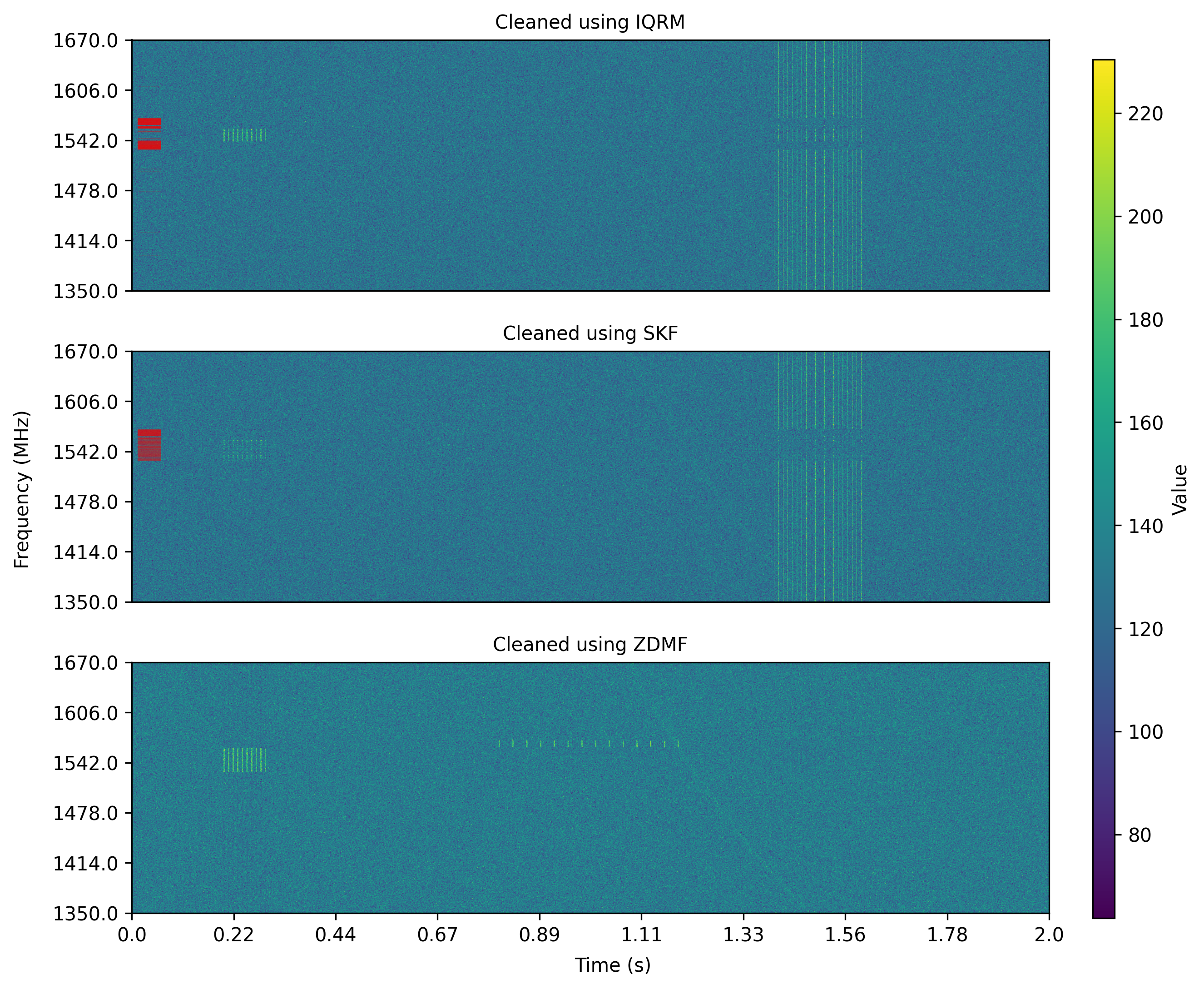}
    \caption{Demonstration of the chosen RFI removal algorithms, run individually, on the data shown in Fig. 1. The red lines at the left of the top two panels indicate the channels that are flagged as RFI-affected by the respective algorithms. \texttt{filtool} (used for SKF and ZDMF) corrects for the bandshape of the chunk of data, whereas \texttt{iqrm-apollo} does not. To present the data comparable to each other, the topmost plot in the figure showing data cleaned by IQRM is, therefore also shown after correcting for its bandshape. Since ZDMF acts on the time samples, it does not mask any data but changes every time sample, and so only removes broadband RFI. One can also notice the patterns extending the residual RFI across all the frequency channels.}
    \label{fig:rfim_on_tvs2}
\end{figure*}

Software implementations of the selected algorithms are installed on the machine that runs the tests. These are capable of reading a filterbank file, cleaning the RFI and writing out a cleaned filterbank file, which is then used to search for the injected pulses. The algorithms being examined are:
\begin{itemize}
    \item Spectral Kurtosis Filter (SKF) \citep[see][]{skf}
    \item Inter Quartile Range Mitigation (IQRM) \citep[see][]{iqrm}
    \item Zero DM Matched Filtering (ZDMF) \citep[see][]{zdmf}
\end{itemize}

\noindent We refer the readers to the papers listed above for the details of the respective algorithms. SKF and ZDMF are implemented in \texttt{filtool}, which is a part of the \texttt{PulsarX}\footnote{https://github.com/ypmen/PulsarX ; not same as \texttt{filtools}} package \citep[see][]{pulsarX} and IQRM is implemented by \texttt{iqrm-apollo}\footnote{https://gitlab.com/kmrajwade/iqrm\_apollo}. The efficacy of the individual RFIM algorithm is shown in Fig.~\ref{fig:rfim_on_tvs2}. 
The flagged data is replaced with noise samples drawn from a normal distribution. In the case of \texttt{filtool}, the mean and standard deviation of the entire data (both flagged and unflagged) is used to generate these noise samples\footnote{In \texttt{filtool}, the data is normalised to have 0 mean and 1 standard deviation, thus using the same mean and standard deviation to replace the data. These values have to be scaled up to values appropriate for the 8-bit data type used in the filterbank file format before exporting.}. However, in \texttt{iqrm-apollo}, the median and standard deviation of the data block are used to generate the distribution of samples to replace the flagged data. \texttt{filtool} and \texttt{iqrm-apollo} are configured to read and clean $2~\mathrm{s}$ (i.e. 31250 time samples) duration subsets of data, and these subsets are used to calculate the statistics and spectral moments (mean, standard deviation, skewness and kurtosis). In a real-time search pipeline, we might not know the nature of the RFI, but in our case, we have prior information about the RFI. This could create a bias in optimising parameters of RFI excision algorithms. To avoid this, we chose to use the default subsets time from \texttt{filtool (PulsarX)} to clean the data. The side effects of this can be seen in Fig. \ref{fig:rfim_on_tvs2}, where some residual RFI is visible, which may not be expected when cleaned using a perfect RFI excision algorithm. However, this reproduces a real observation scenario, where some RFI usually leaks through the filtering stage even after RFI removal.

An important parameter in the SKF and IQRM algorithms is the threshold, which decides whether a slice of data is affected by RFI or not. It is set to $3\mathrm{\sigma}$ as this is a commonly used value for this parameter. In the case of SKF, $\mathrm{\sigma}$ is the inter-quartile range of the skewness and kurtosis values calculated for every frequency channel, whereas for IQRM it is the ratio of the inter-quartile range and inverse cumulative distribution at both quartiles. In the case of IQRM, another parameter needs to be set for optimal RFI rejection. This is the channel radius which is set to approximately $\mathrm{ 10\% }$ of the total number of frequency channels ($410$ in our case). ZDMF has no configurable parameters and instead performs the same action on any data provided to it. Due to specialisations of these algorithms to excise a particular kind of RFI, we use them in groups. However, the performance of each of the mentioned algorithms is heavily dependent on the input provided to it, hence, we do not use every possible permutation of these algorithms. We classify SKF and IQRM to be good at cleaning narrowband periodic RFI and cannot excise RFI that alters noise baselines, such as broadband interference, as described by \citet{iqrm}. This is because the injected broadband RFI does not cause a significant deviation in the per-channel statistics over the time ranges used for calculating the statistics. Therefore, a combination of IQRM - ZDMF or SKF - ZDMF, in theory, should be able to clean most RFI, and this is why TVS-2 was cleaned once with each combination. In addition to this, we also used IQRM alone to clean TVS-2 to get an idea of how important broadband RFI removal techniques are.

This paper focuses only on a limited number of mitigation algorithms. The chosen algorithms are widely used in the fast transient searching community and a systematic comparison of their efficacies will inform their use in future surveys. However, our approach can evaluate any algorithm that can clean dynamic spectra.

\subsection{Single pulse search}
\begin{table}
    \centering
    \caption{Dedispersion plan used for dedispersing time-frequency data.}
    \begin{tabular}{|c|c|c}
    \hline
        Start DM & End DM & DM step \\
        \hline
        0.0 & 100.0 & 0.1 \\
        100.0 & 300.0 & 0.2 \\
        300.0 & 700.0 & 0.4 \\
        700.0 & 1500.0 & 0.8 \\
        1500.0 & 3100.0 & 1.6 \\
        \hline
    \end{tabular}
    \label{tab:ddplan}
\end{table}

\begin{table}
	\centering
	\caption{The parameters of sifting and clustering used in the search.}
	\label{tab:clustering_sifting}
	\begin{tabular}{|l|l|l|} 
		\hline
		Module & Parameter & Values\\
		\hline
		\multirow{3}{*}{Clustering} & Time tolerance & $\mathrm{100~ms}$ \\
                                 & Pulse width tolerance & $\mathrm{100~ms}$ \\
                                 & DM tolerance & $\mathrm{5~pc/cm^3}$ \\ 
        \hline
        \multirow{3}{*}{Sifting} & S/N threshold & $ < 6$ \\
                                 & Pulse width threshold & $ >\mathrm{1~s}$ \\
                                 & DM threshold & $<\mathrm{5~pc/cm^3}$ \\
		\hline
	\end{tabular}
\end{table}

\label{sec:sps_process}
A Python-based testing framework, \texttt{ProTest}\footnote{https://gitlab.com/ska-telescope/pss/ska-pss-protest/-/releases/4.1.2} is used to carry out the task of iteratively running the RFI cleaned test vectors through the search pipeline using the \texttt{ska-pss-cheetah} pipeline framework and verifying the candidate output. As previously described, since we are interested in the influence of RFI mitigation approaches on the detectability of single pulses in an untargeted search, we emulate a real single pulse search. Therefore, the post-RFI excision data is dedispersed at multiple DM trials before being searched for single pulses (shown in Table~\ref{tab:ddplan}). 

Dedispersion is a process of correcting for the delays caused by the interaction of electromagnetic waves with electrons in the Interstellar Medium (ISM) along the line of sight from the telescope to the source of the pulse. This is done by correcting for these delays in each frequency channel and then adding all the frequency channels to increase the S/N of the pulse. These operations are performed by an AVX-512-based tree dedispersion algorithm in an implementation called Klotski\footnote{https://gitlab.com/ska-telescope/pss/ska-pss-cheetah/-/tags/0.4.0} \citep[see][]{klotski_ref}. Emulating real single pulse search process also allows us to see if incomplete RFI removal, for example, results in too many false positive detections\footnote{See appendix \ref{sec:errors} for a description of the detrimental effects of too many candidates on the single pulse search pipeline.} and/or recovery of pulses with incorrect pulse parameters.

The dedispersed time series are formed and convolved with a set of box-car filters of various widths. When the S/N of the convolved product crosses a defined threshold, it is considered a detection. The S/N of a pulse is the integrated power of the pulse, normalised by the square root of the width of the pulse divided by the standard deviation of the off-pulse data. When a detection occurs, the corresponding timestamp of the event, the value of the DM used to dedisperse the data, the width of the box-car and the S/N are recorded. The range of widths of box-car filters begins with 2 samples and increases in powers of 2 up to 8192 samples and then a final filter of 15000 samples (i.e. $\mathrm{960~ms}$) to be a near match to the maximum input pulse width.

The detections are passed through a clustering process to reduce the number of detections corresponding to the same event. We use the friends-of-friends algorithm explained in \citet{FoF_paper}. In this algorithm, detections are grouped based on the proximity of two detections defined by a range of arrival times, DMs, and pulse widths called tolerances. The clustering parameters are listed in Table~\ref{tab:clustering_sifting}, and are set carefully to cluster single pulses of the same events.

The sifting algorithm employed here is a simple method of sifting false detections. A candidate is removed from the list of detections if either DM or S/N are below a defined threshold, or the width is above a threshold. Table~\ref{tab:clustering_sifting} shows the default values of parameters used for sifting. A list of sifted detections is then obtained with corresponding metadata.

\subsection{Recovery of pulse}
\label{sec:pulse_recovery}
The pulse arrival time, DM and width from the list of detections are cross-checked with a list of the expected parameters from the injected pulses which are known for each test vector.  As the input parameters describing the simulated pulses may not be recovered exactly, a range of tolerances is defined for each parameter based on the known value. These tolerances are used to classify if the detected pulse was the one which was originally injected and not a false positive. 

The tolerance on the arrival time of the pulse is taken to be the $1\sigma$ width of the input Gaussian pulse. The tolerance on the width is taken to be one box-car width narrower/wider than the expected pulse width. For example, for a 10-bin-wide input pulse, 8 and 16 bins were also considered to be a detection.

The DM tolerance ($\Delta \mathrm{DM}$) is defined as,
\begin{equation}
\label{equ:dm_tol}
    \Delta \mathrm{DM} = \frac{n \mathrm{w}}{\mathcal{D}(f_{\rm low}^{-2} - f_{\rm high}^{-2})}
\end{equation}
where, $\mathrm{w}$ is the true width of the pulse and $\mathcal{D}$ is the dispersion constant ( $\approx 4.15 ~ \mathrm{GHz^2 ~ cm^3 ~ pc^{-1}~ ms}$). The value of $n$ sets the tolerance to a level with an acceptable recovered S/N when dedispersed at a wrong DM and here $n=2$ allows 85\% of S/N recovery \citep[see][]{sps_technique}. 

Results are summarised in the section below and the fractions of recovery for every test vector are available in appendices \ref{realistic_rfi_appendix} and \ref{severe_rfi_appendix}. It is to be noted that the width of the injected pulse heavily influences the tolerances on arrival time and the DM, hence the tolerances for the case of the widest pulse in the test vector set could lead to defining tolerances so large that it is likely for a false positive to be considered as a true detection. This is the reason why, in some extreme cases, we do not include detections from the widest pulses in the analysis.

We also note that now and again there might be an occasional pulse missed in our analysis, see e.g.~\ref{fig:extreme_iqrm_40ms}, where we would expect to get 6 out of 6 pulses. After inspection, we conclude that this happens when the pipeline returns a time of arrival and/or width that is outside of the aforementioned tolerances. This likely happens as the best S/N occurs at a wider width because of the noise properties and the pulse shape. This reflects the challenges of putting in place methods to check the performance of these pipelines without manual inspection. Such methods are necessary when trying to test across thousands of scenarios that might arise in a real search like those presented here. This loss of an occasional pulse does not affect our interpretation of the algorithms under test, as the overall trends are what we are most interested in. We also note that the pulses would not remain undiscovered in a real search, but the discovered width and possibly DM would differ slightly from the true value. The results for the no-RFI case presented in Appendices \ref{realistic_rfi_appendix} and \ref{severe_rfi_appendix} can be used for comparison with the cases where RFI was included if there is any doubt.

\section{Results}
\label{sec:results}
\subsection{RFI scenario 1}
\label{sec:realistic_rfi}
\begin{figure}
    \centering
    \includegraphics[width=0.8\linewidth]{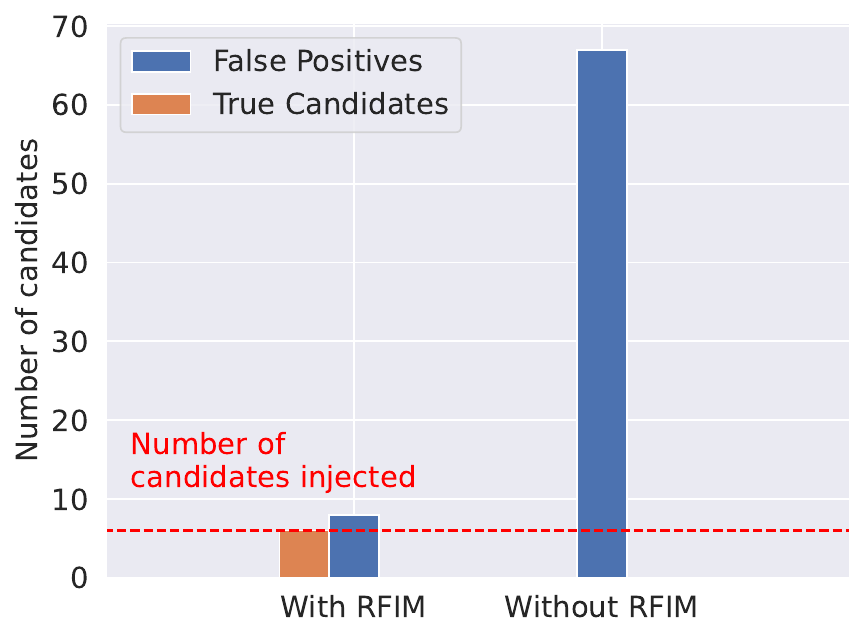}
    \caption{Histogram of the number of true and false positive candidates detected with and without RFIM in a single pulse search. A test vector containing 6 pulses with a DM of $300\,\mathrm{pc/cm^3}$, a width of $\mathrm{40\,ms}$, a single-pulse S/N of $42.4$, and a combination of all three types of RFI ($2\sigma$) is searched for single pulses. Once without RFI cleaning and once after cleaning using IQRM and ZDMF.}
    \label{fig:with_without_rfim}
\end{figure}

\begin{figure*}
    \centering
    \includegraphics[width=0.8\linewidth]{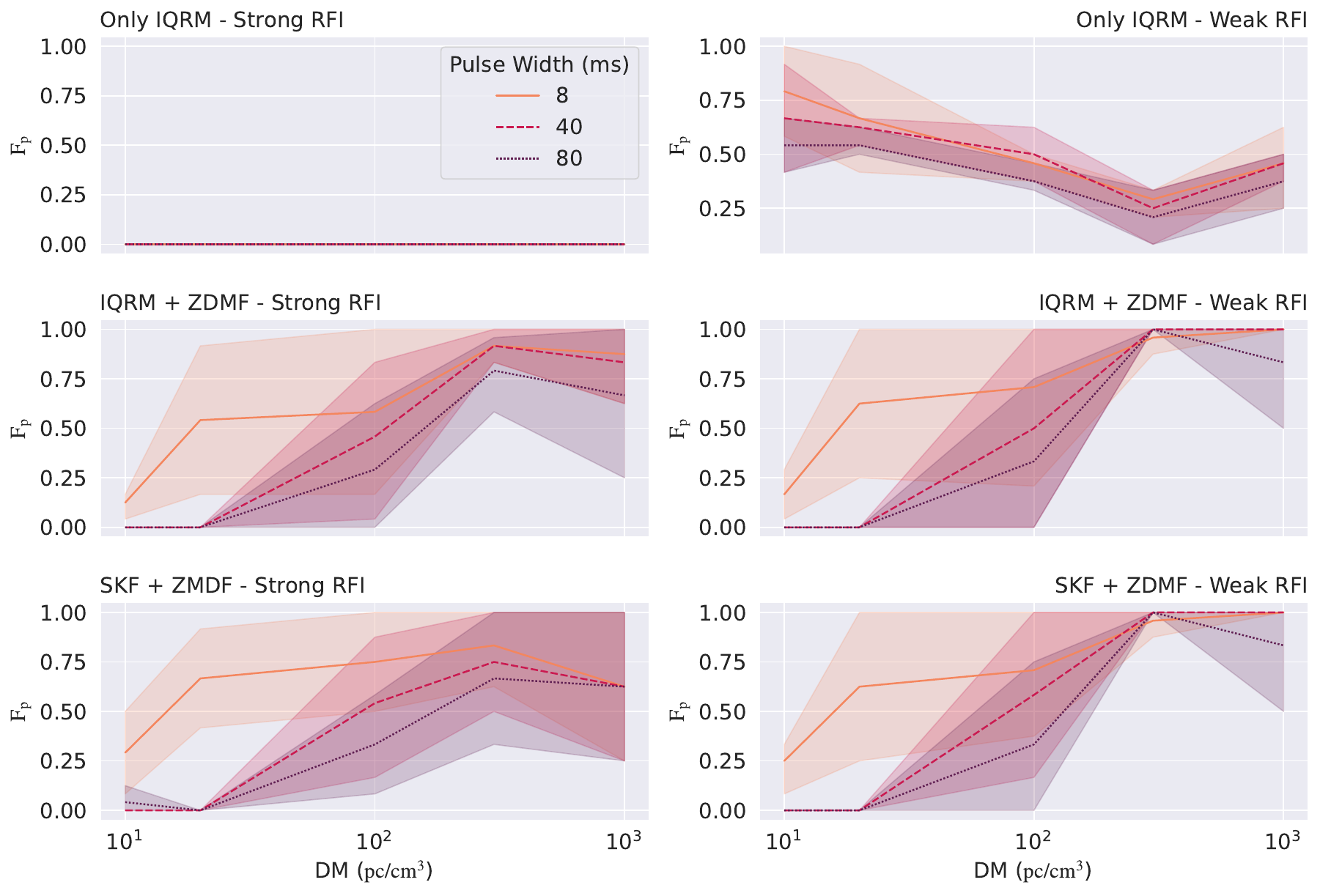}
    \caption{Detection rates for different combinations of RFIM applied to test vectors containing realistic RFI as a function of DM are shown. The fractions of recovered pulses ($\mathrm{F_p}$) are plotted against $\mathrm{DM}$ for multiple widths. The combinations of RFIM used are - IQRM, IQRM + ZDMF and SKF + ZDMF. The plots on the left side are obtained from results when test vectors from TVS-2 containing RFI of type 8 are cleaned with respective RFIM, and the ones on the right side contain RFI of type 1 (see Table~\ref{tab:realistic_rfi_testvectors}). The shaded region indicates the range of $\mathrm{F_p}$ for different S/N (ranging from 14.1, corresponding to the lower edge, to 141.4, corresponding to the upper edge) at a given DM for a given pulse width. In the first panel, where the test vectors containing RFI of type 8 are cleaned using IQRM, no pulses are recovered, but the fraction of recovery increases when the RFI is weaker and/or cleaned by ZDMF as well.}
    \label{fig:real_rfi_all_rfim}
\end{figure*}

The use of RFIM has advantages and disadvantages. Although RFI mitigation algorithms may be effective at removing interference, they may affect data quality, leading to unintended distortion or weakening of the signal of interest. Many statistical parameters may also change, there might be an observed reduction in S/N of the pulse and sometimes cleaning techniques remove parts or the whole of the pulse if it is bright enough. In addition to this, it costs additional computing power and time. However, in the presence of RFI, the observed data cannot be used for most science cases because RFI might interfere with the process of extracting astrophysically useful and accurate information from it.

As an example to further demonstrate the need to use RFIM, a test vector containing pulses of width $\mathrm{40~ms}$ and a DM of 300 $\mathrm{pc/cm^3}$ and added RFI was searched without cleaning by any RFI excision method. Fig.~\ref{fig:with_without_rfim} shows the number of true and false positive detections. The same test vector was cleaned using IQRM followed by the ZDMF algorithm, before searching for single pulses and we could recover all 6 pulses. From Fig.~\ref{fig:with_without_rfim}, we can infer that, although there are a significant number of false positives, we could still detect injected pulses, contrary to the case of not using RFIM, where none of the pulses are detected. Therefore, as expected, it becomes clear that using RFIM during the search for transients is essential.

We used a combination of RFIM algorithms to clean the contaminated test vectors from TVS-2 as well as a complementary set with no RFI injected for comparison. The test vectors were cleaned using IQRM, IQRM with ZDMF, and SKF with ZDMF. In each case the resultant filterbank files were searched for single pulses and the results were analysed. Ideally, we should recover all the pulses for all the DMs, S/Ns and pulses of all widths. Fig. \ref{fig:real_rfi_all_rfim} shows the fraction of recovered pulses using all three RFI mitigation strategies on test vectors from TVS-2 with strong and weak RFI of all types (Type 1 and Type 8 from Table \ref{tab:realistic_rfi_testvectors}). These are representative numbers from just 2 realisations of RFI, more results on the entire set, including those with no RFI, can be found in appendix \ref{realistic_rfi_appendix}\footnote{We note that for a handful of test vectors with no RFI injected (both in Appendices \ref{realistic_rfi_appendix} and \ref{severe_rfi_appendix}), the pipeline returned zero candidates when the S/N was very high. We identified a corner case in our pipeline where this is because a lower-DM reports a higher S/N. In these cases, we have instead reported the results after manual inspection. We have also confirmed that this does not affect any of the other results.}. In the first panel of Fig.~\ref{fig:real_rfi_all_rfim}, the fraction of recovered pulses is zero because the search pipeline timed out in all the cases. This is due to residual RFI creating so many candidates that the search pipeline stalls and is eventually killed by \texttt{ProTest}, leading to no pulses being recovered (see Appendix \ref{sec:errors} for more details).

Our simulations included test vectors with widths of 800~ms, and the results for those widths are presented in the Appendices \ref{realistic_rfi_appendix}. However, as briefly mentioned above, in the presence of incompletely removed RFI, the wide box car needed to recover these wide pulses can result in low-S/N false positive detections because of the larger tolerances on DM, width and arrival time. We therefore do not include it in further analysis. However, we do present these results in the appendices to highlight the challenges that single pulse searches in the time domain will have in detecting such wide pulses in the presence of RFI. This also suggests that other approaches to RFI removal might be needed to enable their detection.

In the presence of all three kinds of RFI, we fail to recover a large portion of the injected pulses from the single pulse search when IQRM alone is used to clean the RFI. This is especially true in the presence of strong levels of RFI of all types ($2\sigma$), where we recover the fewest pulses. The fraction of recovered pulses increases when ZDMF is used in sequence with IQRM. This shows that IQRM is not good at cleaning broadband RFI and should always be used in combination with an RFI excision method that is good at removing broadband RFI (as explained by \citet{iqrm}). Another observation is that when ZDMF is used in sequence with SKF or IQRM, either combination of RFIM algorithms works well, and the total number of pulses recovered increases when compared to the case of not using ZDMF, but the fraction of recovered pulses was not $1$. We could however, recover more pulses at higher DMs when compared to the fraction recovered at lower DMs. It appeared as if there was a function governing the process due to which pulses were not detected. To gain a better understanding of what could be causing this, we decided to use RFI-free test vectors containing pulses and apply the different mitigation methods individually.
\subsection{RFI scenario 2}
\label{sec:no_rfi}
\begin{figure}
    \centering
    \includegraphics[width=0.85\linewidth]{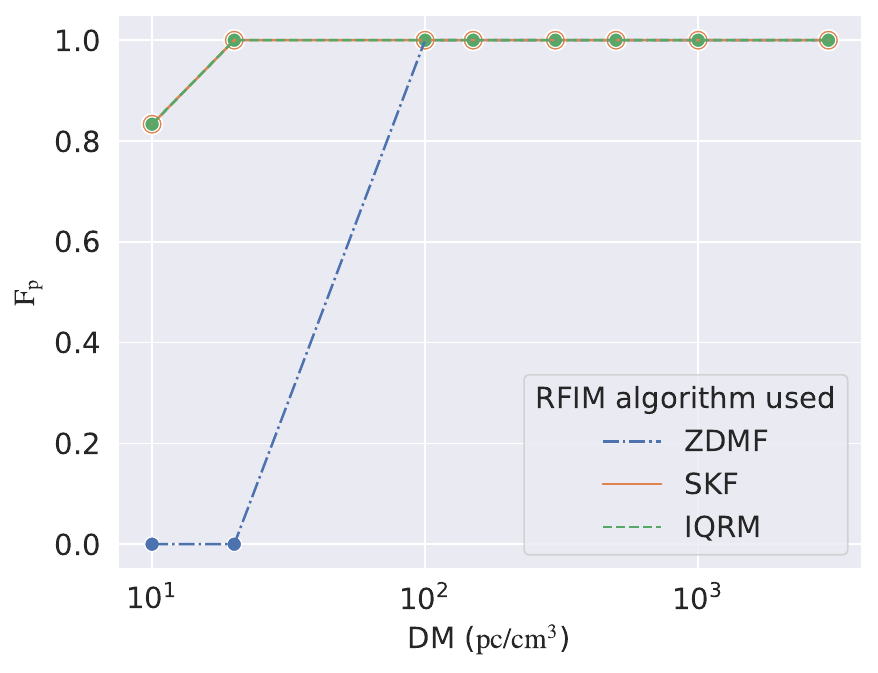}
    \caption{The impact of each algorithm on the fraction of recovered pulses ($\mathrm{F_p}$) across a range of  DMs for test vectors without added RFI. The test vectors contain pulses with a width of $\mathrm{40,ms}$ and a (S/N) of 42.4. Note that the $\mathrm{F_p}$ curves for SKF and IQRM overlap in the figure.}
    \label{fig:rfim_on_no_rfi}
\end{figure}

\begin{figure*}
    \centering
    \includegraphics[width=0.75\linewidth]{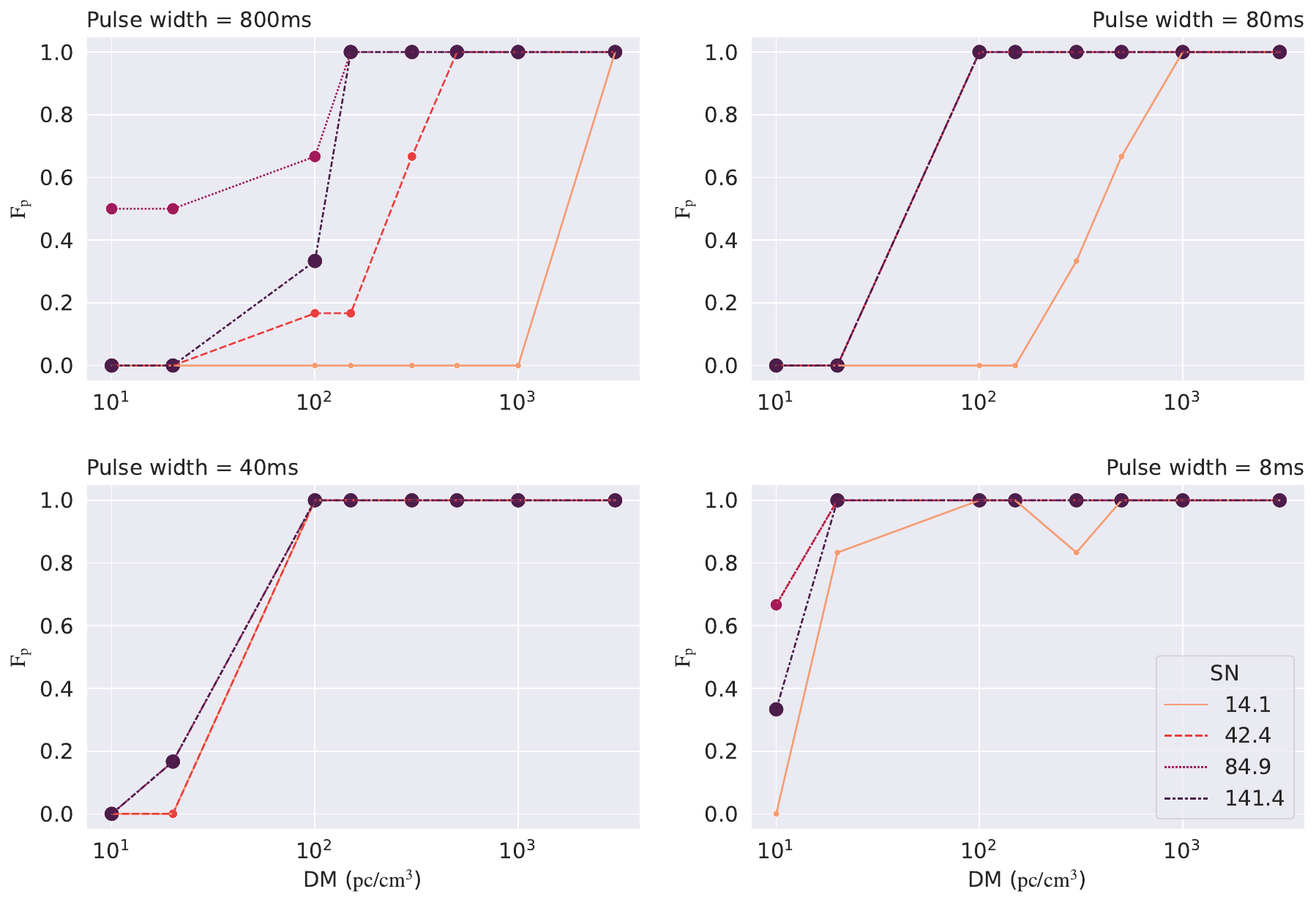}
    \caption{The effect of ZDMF filtering on RFI-free test vectors containing pulses of various pulse widths and S/N is shown by plotting the fraction of recovery of pulses ($\mathrm{F_p}$) as a function of DM. At low DM, it becomes more difficult to recover wider pulses, however, this difficulty diminishes as the S/N increases. We have included the 800~ms widths here as no RFI was injected (see discussion in section \ref{sec:realistic_rfi}).}
    \label{fig:zdot_all_widths}
\end{figure*}

To examine the cause of the failure to detect some pulses, each of the RFI mitigation algorithms was applied to test vector data free of RFI contamination (TVS-1). Fig. \ref{fig:rfim_on_no_rfi} shows an example of test vectors containing pulses of width $\mathrm{40\,ms}$ and a S/N of $\approx42$, where the fraction of pulses recovered from the single pulse search are plotted as a function of the DM of the pulses (the full results can be found in the left most panel of the figures in appendix \ref{realistic_rfi_appendix} and \ref{severe_rfi_appendix}).
We observed that, in our sample space, the single pulse search did not recover pulses with a DM below $100~\mathrm{\,pc / cm^3}$ in all scenarios that included ZDMF filtering. With SKF and IQRM on their own, the fraction of recovered pulses is almost complete for all the DM trials when either of the algorithms is used to clean the RFI-free test vectors (see Fig. \ref{fig:rfim_on_no_rfi}).

To investigate the lack of low-DM detections in the ZDMF case, we looked at the results for all the test vectors cleaned by ZDMF as shown in Fig. \ref{fig:zdot_all_widths}. We see a clear loss of pulses at low-DM when ZDMF is used. It is to be noted that we did not generate test vectors that contain pulses with DMs between $20 - 100~\mathrm{\,pc / cm^3}$ as our intention was not to test ZDMF alone. We therefore do not know exactly where this transition might happen, and as we can see in Fig.~\ref{fig:zdot_all_widths}, there is a dependence on pulse width and S/N. This loss of low-DM pulses was already identified as an issue by \citet{zdf} when zero-DM filtering (ZDF) was first introduced, and is due to the influence of the dispersed pulse on the estimation of the mean. 
The ZDMF algorithm is an extended version of the ZDF algorithm, which subtracts only the corresponding contribution from each channel at a given time sample \citep[see][]{zdmf}. As described by \citet{zdf}, the smaller the fractional bandwidth (the ratio of the bandwidth to the central frequency), the greater the prominence of the effect. The fractional bandwidth that we are using for this paper is similar to that for band-2 of SKA-Mid, and thus our results are representative of what can be expected in that band.

In the first panel of Fig. \ref{fig:zdot_all_widths}, the number of recovered pulses at lower DMs is greater for pulses with S/N of 85 than for pulses with S/N of 140. Upon investigating individual pulses with different DMs, we conclude that the detections reported by the search pipeline for an input S/N of 85 were due to the residual of the pulse after cleaning using the ZDMF algorithm. The candidates have S/N of approximately 6 or 7, and their corresponding pulse residuals had barely managed to cross the threshold to be detected by the search pipeline.

\subsection{RFI scenario 3}
\label{sec:extreme:rfi}
\begin{figure*}
    \centering
    \includegraphics[width=0.75\linewidth]{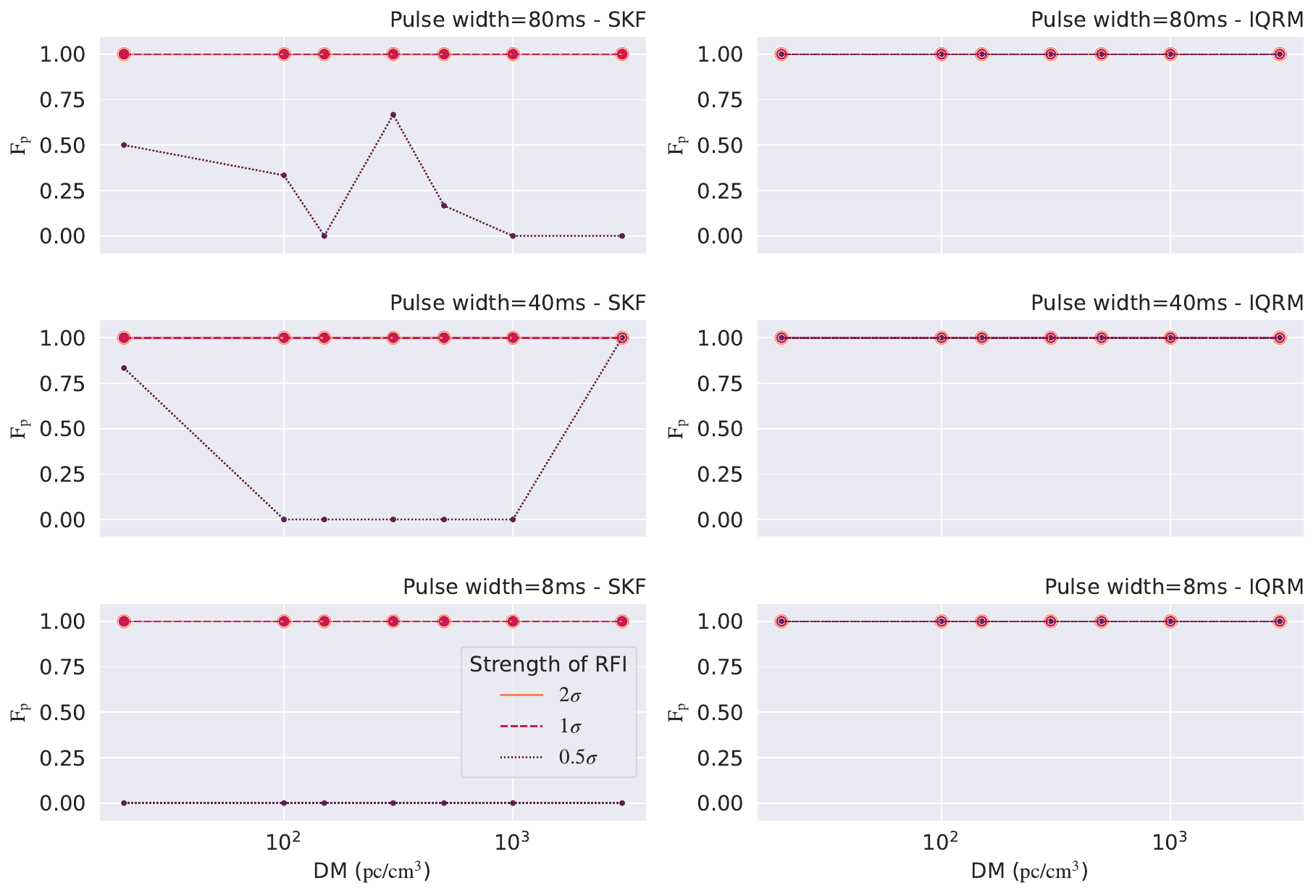}
    \caption{Recovered fractions of single pulses ($\mathrm{F_p}$) as a function of DM for the case of S/N $\approx 85$ for multiple widths when IQRM and SKF algorithms are used for various strengths of narrowband periodic RFI. Note that the red and orange dashes follow each other in all cases and for the plots showing $\mathrm{F_p}$ for IQRM, all three lines follow each other and the value is unity at every DM.}
    \label{fig:iqrm_vs_skf}
\end{figure*}

\begin{figure}
    \centering
    \includegraphics[width=0.75\linewidth]{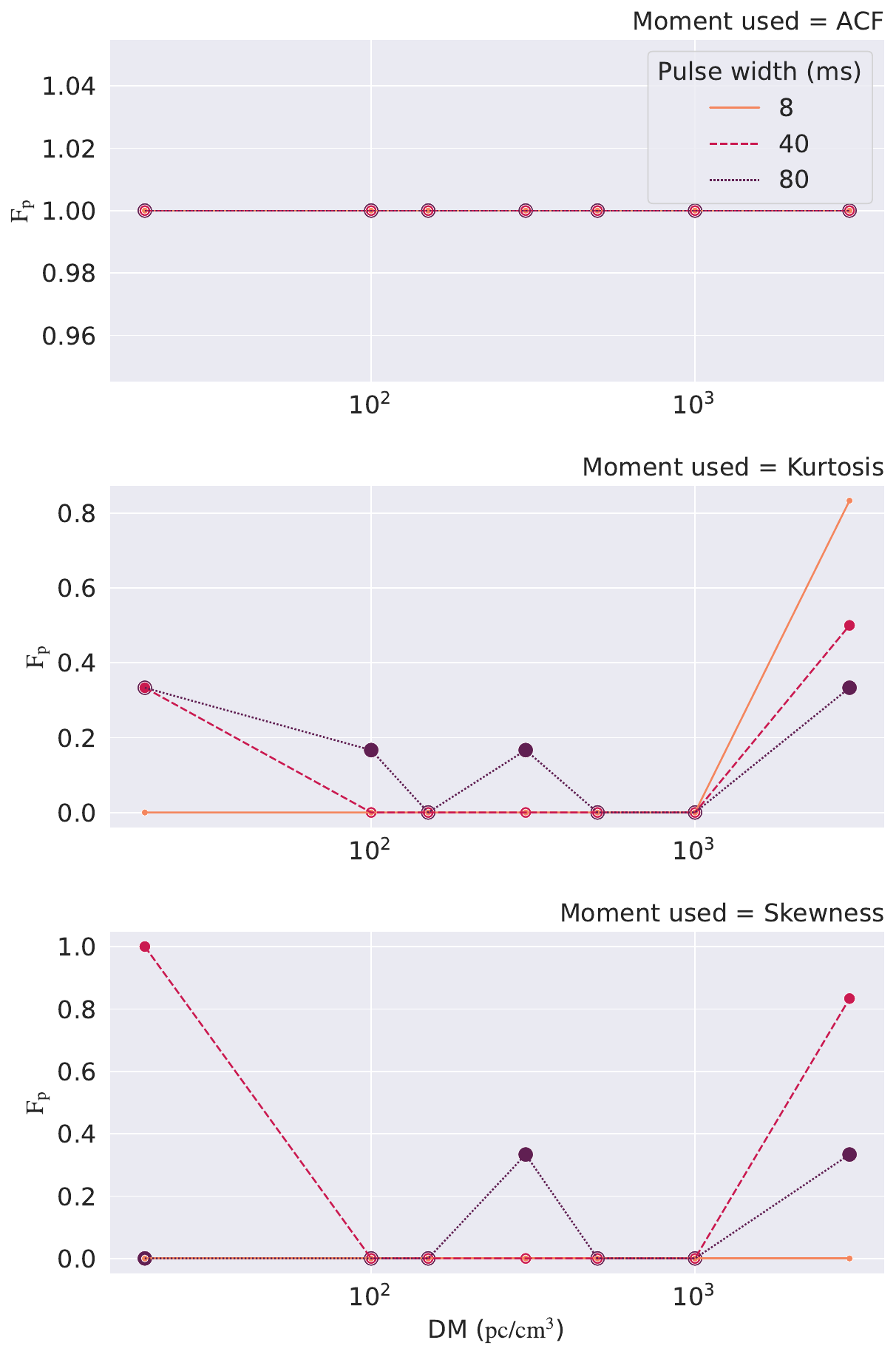}
    \caption{The effect on the fraction of recovered pulse ($\mathrm{F_p}$) at a range of DMs for using various moments for RFI excision by IQRM is shown in the figure. The test vectors used contain periodic narrowband RFI of strength $0.5\sigma$ and pulses with S/N $\approx 42$. Note that in the plot showing $\mathrm{F_p}$ vs DM when the autocorrelation factor (ACF1) spectral metric is used, all three lines (orange, red and purple line) follow each other for all the values of DM.}
    \label{fig:iqrm_various_moments}
\end{figure}

\begin{figure*}
    \centering
    \includegraphics[width=0.75\linewidth]{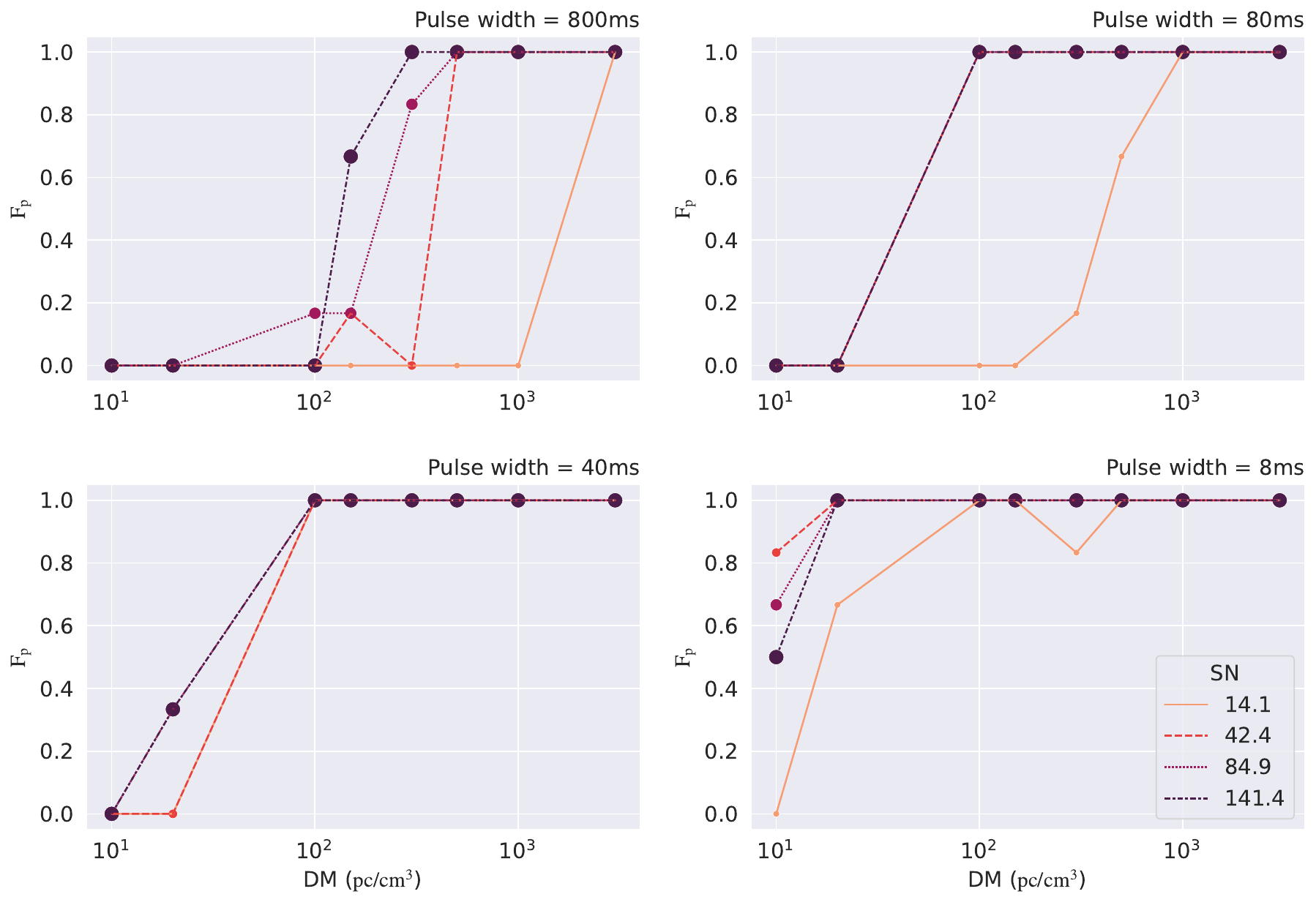}
    \caption{The fraction of recovery of single pulses ($\mathrm{F_p}$) as a function of DM for multiple widths in an RFI environment which has a $2\sigma$ level of broadband RFI affecting 10\% of total time samples observed. The result is similar to Fig. \ref{fig:zdot_all_widths} which indicates that the ZDMF algorithm is unaffected by the amount of broadband RFI in the data.}
    \label{fig:zdot_rfi_for_all_width}
\end{figure*}

\subsubsection*{Excision of Narrowband periodic RFI}
\label{sec:narrowband_extreme}
A set of test vectors containing narrowband periodic RFI (from TVS-3B, see Table \ref{tab:extreme_rfi_testvectors}) is was cleaned using both SKF and IQRM (see section \ref{sec:rfim_details}), and the results can be found in appendix \ref{severe_rfi_appendix}. Fig. \ref{fig:iqrm_vs_skf} shows the effect of DM on the recovery of pulses with S/N $\mathrm{\approx\,85}$ and a range of widths. In both cases, when the data is cleaned by either IQRM or SKF, there is a complete recovery of single pulse events in an environment containing RFI of strength $\mathrm{\geq 1\sigma}$, where $\mathrm{\sigma}$ is the RMS of the noise. However, when the strength of the injected RFI is fairly low (here $0.5\mathrm{\sigma}$ per frequency channel per time sample, which is low in our test vector sets, because, when individual time samples or frequency channels are considered, the RFI is indistinguishable from the noise), IQRM can mitigate the RFI well enough that all the pulses are recovered, but SKF leaves behind residual RFI due to which some pulses cannot be detected during the single pulse search. This can be seen in the left panels of Fig. \ref{fig:iqrm_vs_skf}, where we can get zero candidates in the two ways described in Appendix~\ref{sec:errors}. We note that in the middle-left panel of Figure \ref{fig:iqrm_vs_skf} there are some pulses detected at the lowest and highest DMs. In the former case this corresponds to the pulses being detected, although at an incorrect DM, but which is still within the tolerances, while the latter case corresponds to low-S/N residual RFI appearing within the tolerances. It is to be noted that the spectral metric used by IQRM is the ACF1, but SKF uses skewness and kurtosis, which are statistical measures for the shape of a probability distribution (here, the data in a filterbank file), as its spectral moments to flag RFI-affected channels. When IQRM is run with the chosen spectral moment set to be kurtosis and skewness, the fraction of pulses recovered is no longer complete, as shown in Fig. \ref{fig:iqrm_various_moments}. This is consistent with the SKF results and shows that the spectral moment used for the mitigation process plays a vital role in mitigating low levels of RFI, as also discussed in \citet{iqrm}.

\subsubsection*{Excision of Broadband RFI}
As described in section \ref{sec:rfim_details}, broadband interference alters the noise baseline, and the SKF and IQRM algorithms are specialised to remain unaffected by such baseline variations. All the test vectors containing broadband RFI (TVS-3A, see Table~\ref{tab:extreme_rfi_testvectors}) were therefore cleaned only using ZDMF, and the results can be found in appendix \ref{severe_rfi_appendix}. Fig. \ref{fig:zdot_rfi_for_all_width} demonstrates the response of ZDMF by showing the fraction of recovered pulses as a function of DM for all widths in an environment containing RFI of strength equal to $2\sigma$ and affecting 10\% of time samples. The trends in the plot are similar to those demonstrated in the earlier section under no RFI conditions (see Fig. \ref{fig:zdot_all_widths}). The fraction of recovery increases as the pulse width decreases, and increases as the DM increases, but the RFI environment does not show any trend. This is because the operations performed by ZDMF rely on the undispersed nature of RFI rather than its strength. Another interesting observation is that, in the case of high DM but smaller S/N, the recovery of pulses is inversely proportional to the width of the injected pulse. For a given fractional bandwidth, the broader the pulse, the greater will be the degradation in S/N thus suppressing it enough not to cross the detection thresholds (mentioned earlier in section \ref{sec:no_rfi}).

\section{Discussion}
\begin{figure}
    \centering
    \includegraphics[width=0.8\linewidth]{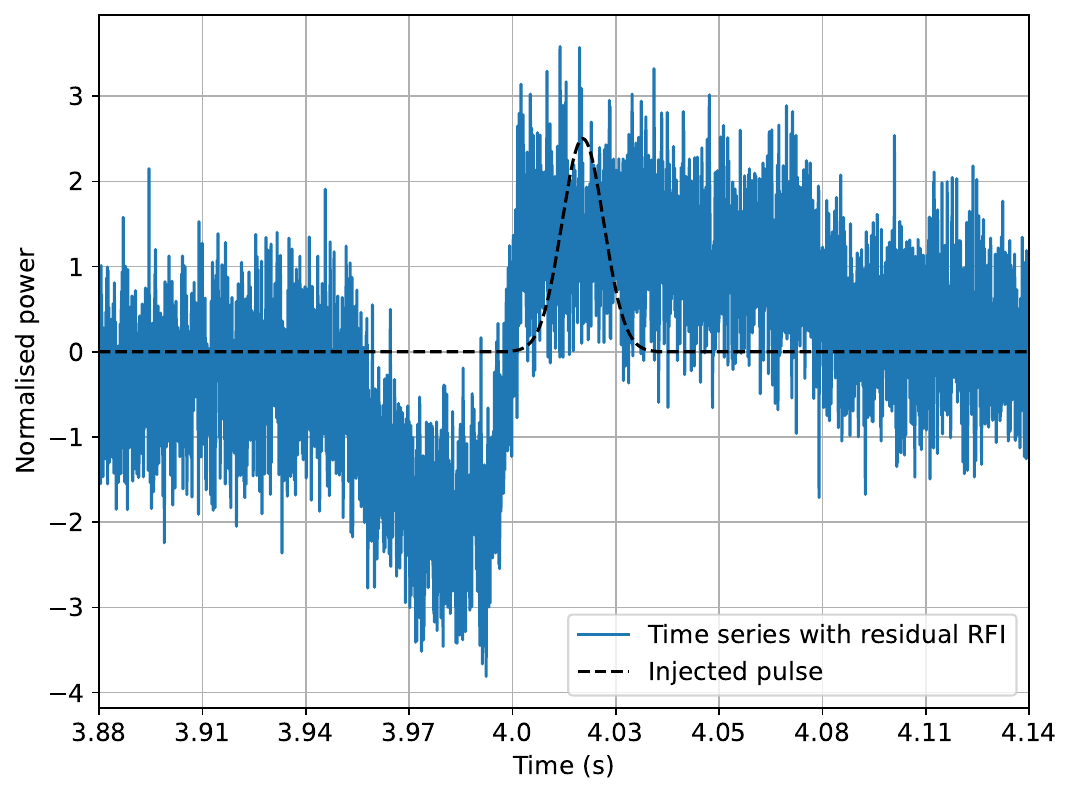}
    \caption{Dedispered time series of a test vector where a pulse with a width of $40~\mathrm{ms}$, S/N of 14.1, and DM 10 $\mathrm{pc/cm^3}$ has been injected. Narrowband RFI of $0.5\sigma$ affecting $10\%$ of frequency channels (see Table \ref{tab:extreme_rfi_testvectors}) was injected and cleaned using SKF. The dashed black line is where the pulse was expected to be but is obscured by the residual RFI.}
    \label{fig:zero_detection_cands_tim}
\end{figure}

As discussed above, our aim was to generate a range of RFI scenarios which included simulated RFI with properties that were drawn from the underlying properties of real-world RFI, namely broadband, narrowband and then periodic versions of both of these, to test some common RFI mitigation algorithms. The highly variable and varied nature of RFI means that even though we considered a large number of different scenarios, it isn't possible to capture all possibilities here. We note that future work should consider the inclusion of more random variability of the amplitude of the RFI when it is injected, including an increase in the noise for example. The nature of the periodic RFI we introduced, especially in the narrowband case, was challenging, but when it was strong, both SKF and IQRM were able to deal with it well, but when weaker, SKF struggled more. Future work should also include a wider range of timescales for the periodic RFI, and compare the performance with the timescales used in calculations of the relevant statistics for each tested RFIM algorithm. Although not directly related to the RFI algorithms under consideration here, the issue of the large number of false-positives that can be produced in a single pulse search if the RFI is particularly bad and thus cannot be adequately removed, needs consideration. This is true whether you are doing an online or offline search, as it can result in your search pipeline stalling. In the real-time search, it may result in you missing real signals, and/or not being able to trigger on them.

Some cases with 0 detections were investigated to determine the cause, especially those cleaned by SKF (see Fig. \ref{fig:extreme_skf_40ms}). The test vectors were searched over a smaller number of DM trials, and the output candidate file from the search pipeline was manually inspected. Fig. \ref{fig:zero_detection_cands_tim} shows an example of a dedispersed time series of a test vector containing a pulse, dispersed with a DM of 10 $\mathrm{pc/cm^3}$, a width of 40 ms, and a S/N of 14.1, along with narrowband RFI of $0.5\sigma$ affecting $10\%$ of frequency channels (refer Table \ref{tab:extreme_rfi_testvectors}). The plot also contains a dashed line, which is where the pulse would be expected to be found. Instead, the residual RFI has led to a detection of a false positive. As the parameters of this false positive lie outside of the tolerances for the real pulse, it is rightly not counted. This is evidence that residual RFI can corrupt pulses, and it can lead to detecting candidates with incorrect pulse parameters, as stated in the introduction.

\section{Conclusions}

\label{sec:conclusion}
We present a method to evaluate the effectiveness of an RFI removal technique by defining a number of test cases that one wants to test for any RFIM algorithm capable of cleaning dynamic spectra. This method is demonstrated using combinations of three algorithms: IQRM, SKF, and ZDMF. Testing of the algorithm independently confirms that a single algorithm is insufficient. We find that across the range of strengths of RFI investigated here, the combination of IQRM and ZDMF works best when the ACF1 spectral metric is used for IQRM. As expected, the use of ZDMF does have a negative impact on the recovery of low-DM pulses for the fractional bandwidth used here. In the future, the investigation of other robust RFIM techniques (including spatial filtering, frequency rejection, and those which use deep learning) using the proposed method for radio transient search is strongly encouraged. Studying the efficacy of algorithms for wider and/or finer sampling ranges for all the parameters that were assumed to be constant or were restricted (such as spectral index and temporal features of pulses, limited realisations of RFI instances) is also encouraged.

\section*{Acknowledgements}

The authors acknowledge funding from the United Kingdom’s Research and Innovation (UKRI) Science and Technology Facilities Council (STFC), project reference [ST/Z510439/1, ST/Z510440/1 and ST/W001950/1]. They would like to thank Yunpeng Men for discussions regarding the ZDMF algorithm, Michael Keith for supporting RFI injection with \texttt{filtools}, Sergio Belmonte D\'{i}az for the discussions on the S/N for single pulses and Aristeidis Noutsos for their comments. The authors would also like to thank the reviewers of the paper, whose comments significantly improved the manuscript.

\section*{Data Availability}

Steps to reproduce the test vectors and the results have been described in the paper along with all the software used.



\bibliographystyle{rasti}
\bibliography{ref} 



\appendix
\section{Investigating the cases of missing candidates and errors}
\label{sec:errors}
\begin{figure}
    \centering
    \includegraphics[width=0.8\linewidth]{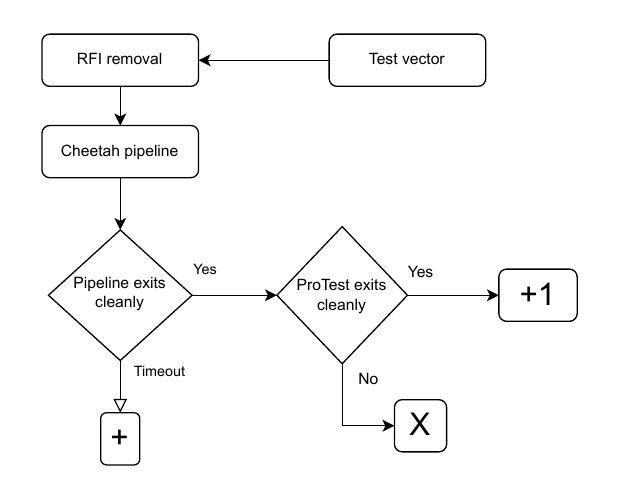}
    \caption{Flowchart describing the single pulse search and validation process. There are 3 results of a pipeline run - The cheetah pipeline may timeout due to a huge number of candidates, denoted by $+$, or the pipeline may run fine, but the exported file maybe corrupted due to a recognised pipeline bug, denoted by \ding{53}, or cleanly exit resulting in a number of recovered pulses.}
    \label{fig:process_flowchart}
\end{figure}

We have identified two scenarios where the pipeline falsely reports no candidates (see Fig.~\ref{fig:process_flowchart}). The first occurs when the pipeline cleanly exits, but the output file containing the information on the candidates gets corrupted. This happens when there are large numbers of candidates and many asynchronous processes are writing simultaneously. This leads to \texttt{ProTest} not being able to ingest the file. Such cases are marked with \ding{53} in the plots in the Appendices. The second scenario occurs when a large number of candidates are produced by \texttt{ska-pss-cheetah} and the clustering algorithm takes too long. This leads to a timeout in the pipeline. These cases are marked with $+$ in the plots in the Appendices. Both cases were revisited and searched over a smaller DM range and all candidates were recovered, supporting the arguments stated in the paper. Both aforementioned cases cause the pipeline to fail, primarily due to a large number of candidates, mainly due to residual RFI, which is the problem.

\section{RFI scenario 2}

\label{realistic_rfi_appendix}
\begin{figure*}
    \centering
    \includegraphics[width=0.9\linewidth]{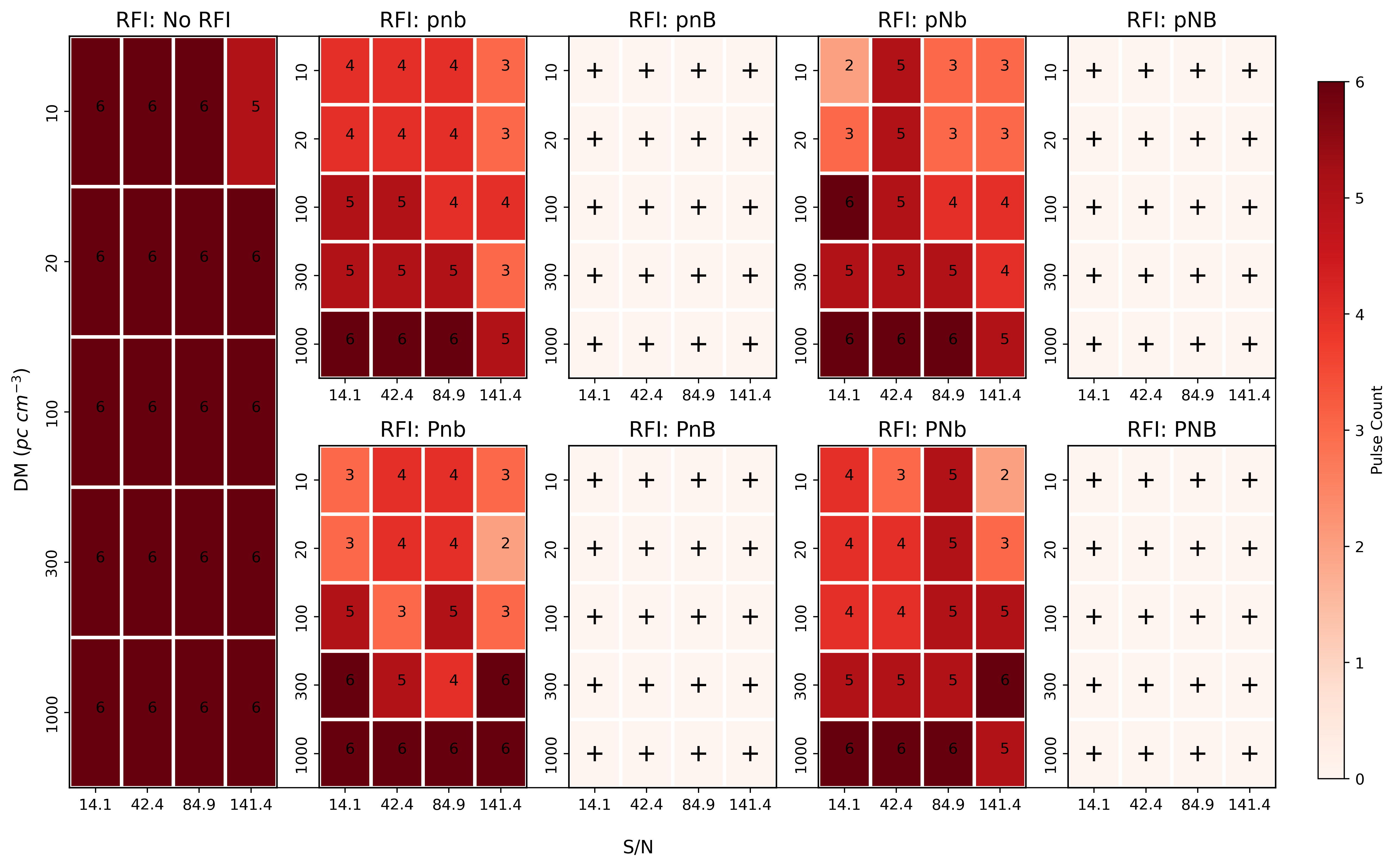}
    \caption{Number of detections of the pulses with a width of 800 ms cleaned using IQRM as a function of DM and S/N. The left-most panel shows the no RFI cases while other plots are for all the RFI combinations given in Table~\ref{tab:realistic_rfi_testvectors}, with small letters (e.g. p) indicating weak RFI and capital letters (e.g. P) indicating strong RFI of that type. See Appendix \ref{sec:errors} for explanation of the $+$'s and \ding{53}'s}
    \label{fig:realistic_iqrm_800ms_mat_plot}
\end{figure*}

\begin{figure*}
    \centering
    \includegraphics[width=0.9\linewidth]{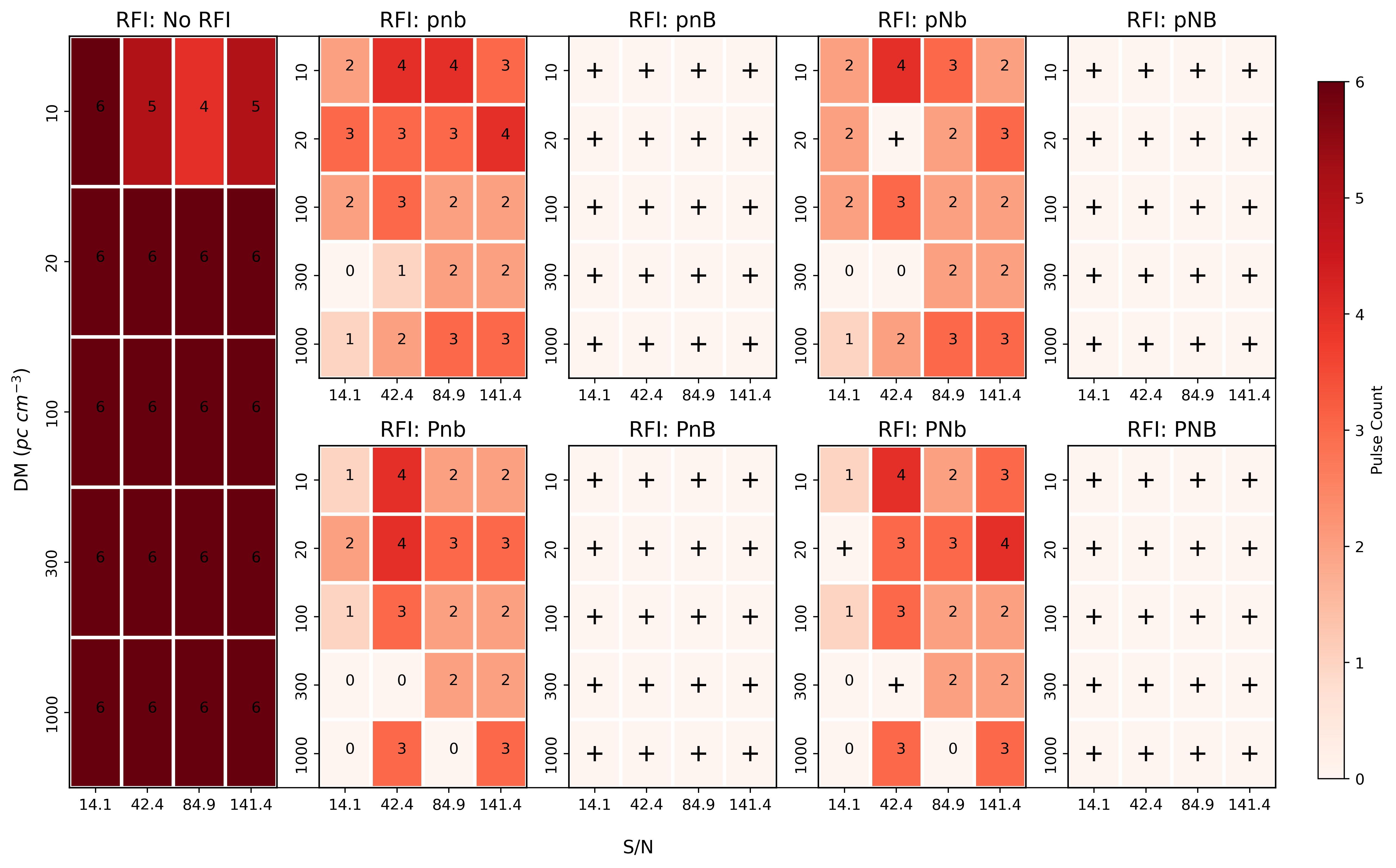}
    \caption{As for \ref{fig:realistic_iqrm_800ms_mat_plot} but for a pulse width 80 ms.}
    \label{fig:realistic_iqrm_80ms_mat_plot}
\end{figure*}

\begin{figure*}
    \centering
    \includegraphics[width=0.9\linewidth]{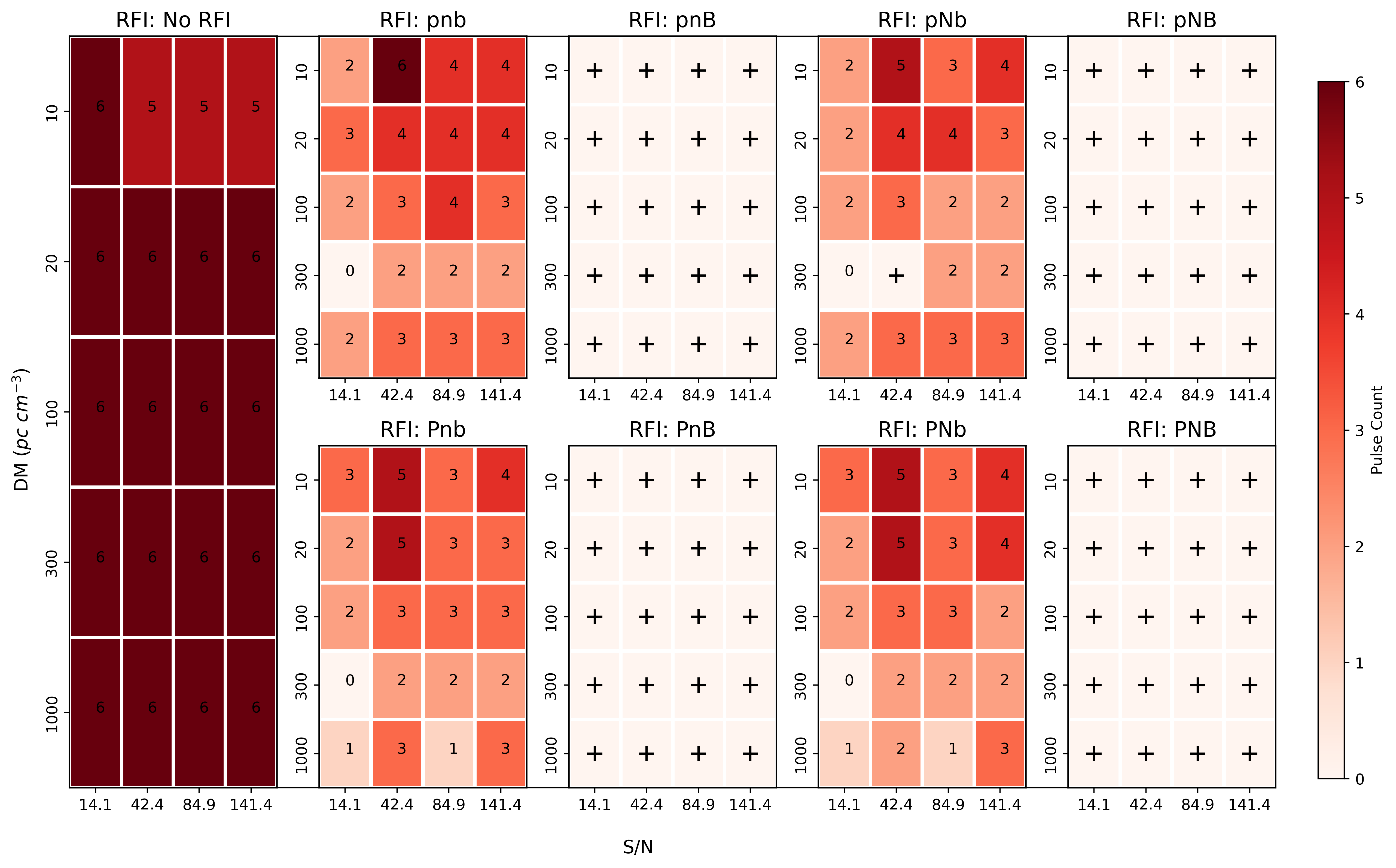}
    \caption{As for \ref{fig:realistic_iqrm_800ms_mat_plot} but for a pulse width 40 ms.}
    \label{fig:realistic_iqrm_40ms_mat_plot}
\end{figure*}

\begin{figure*}
    \centering
    \includegraphics[width=0.9\linewidth]{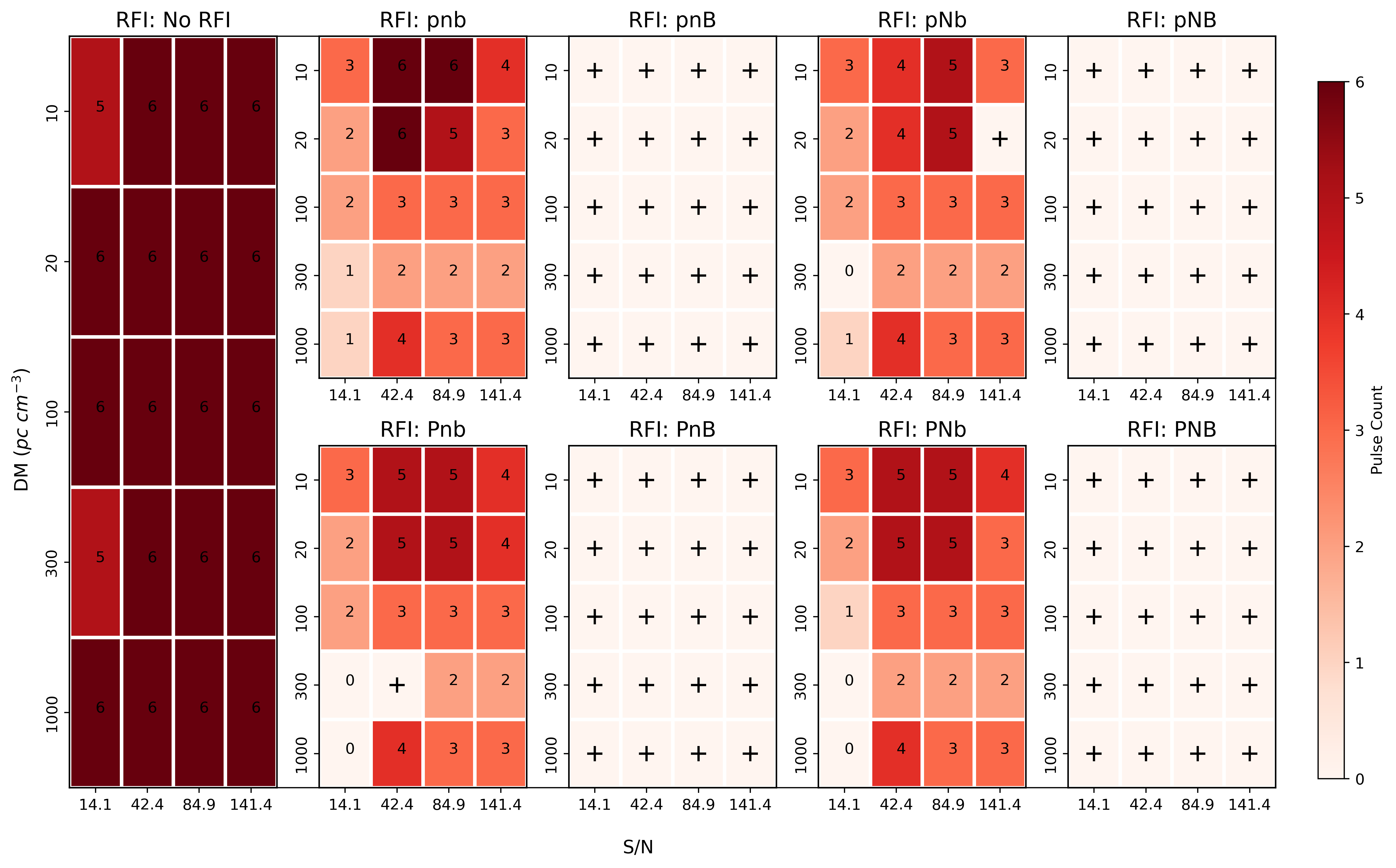}
    \caption{As for \ref{fig:realistic_iqrm_800ms_mat_plot} but for a pulse width 8 ms.}
    \label{fig:realistic_iqrm_8ms_mat_plot}
\end{figure*}

\begin{figure*}
    \centering
    \includegraphics[width=0.9\linewidth]{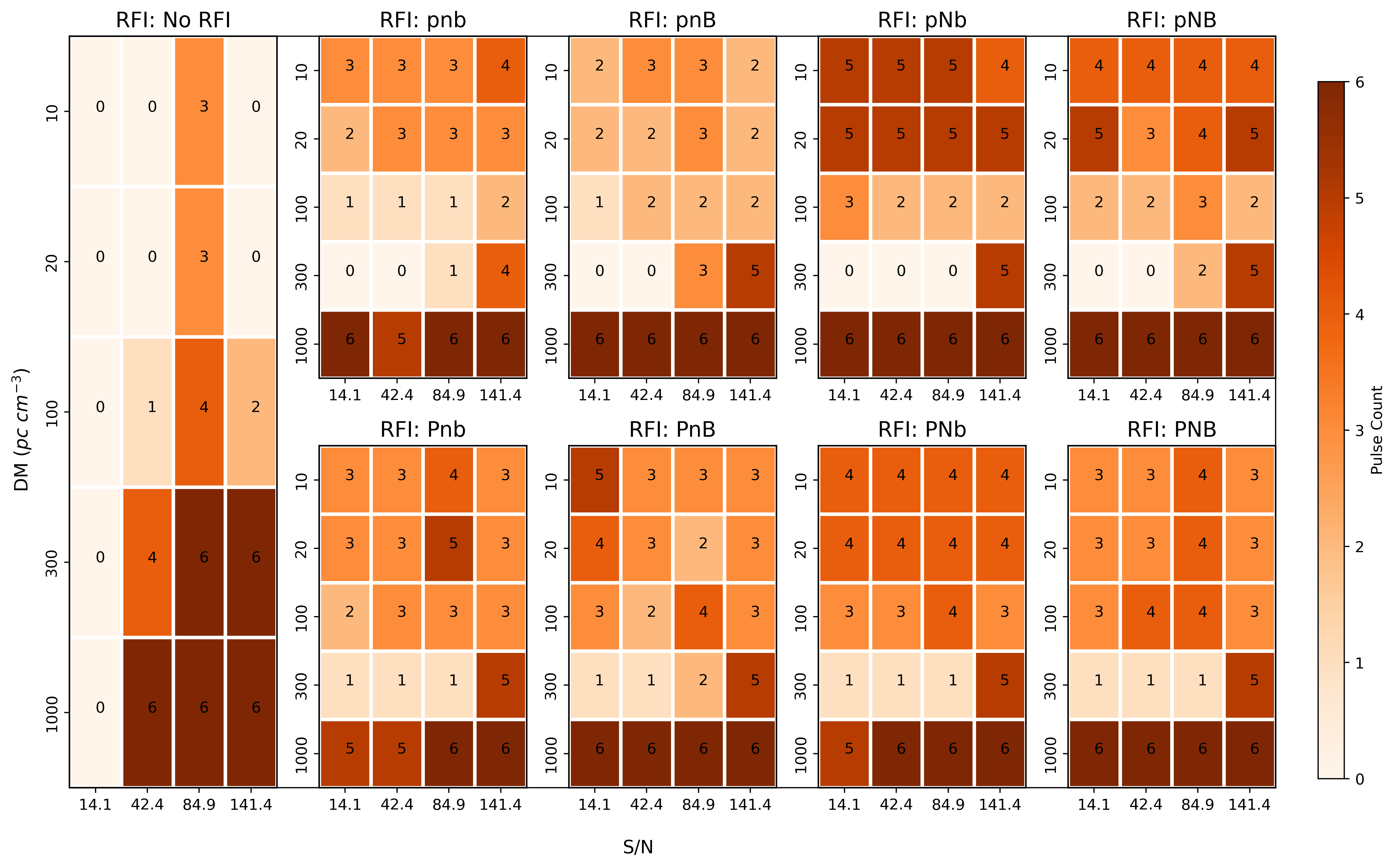}
    \caption{As for \ref{fig:realistic_iqrm_800ms_mat_plot} but IQRM and ZDMF were used, and for a pulse width of 800 ms.}
    \label{fig:realistic_iqrm_zdot_800ms_mat_plot}
\end{figure*}

\begin{figure*}
    \centering
    \includegraphics[width=0.9\linewidth]{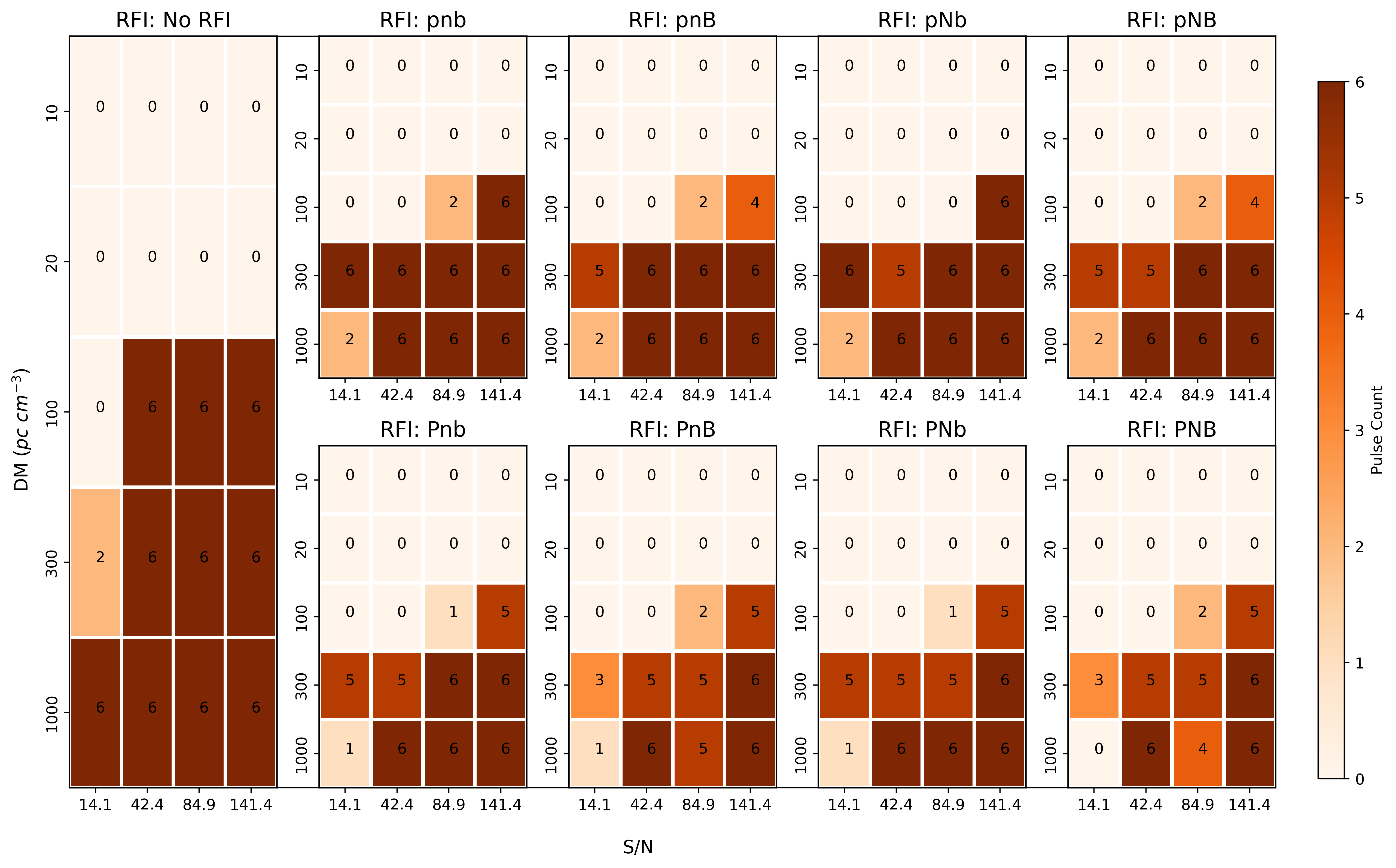}
    \caption{As for \ref{fig:realistic_iqrm_zdot_800ms_mat_plot} but for a pulse width 80 ms.}
    \label{fig:realistic_iqrm_zdot_80ms_mat_plot}
\end{figure*}

\begin{figure*}
    \centering
    \includegraphics[width=0.9\linewidth]{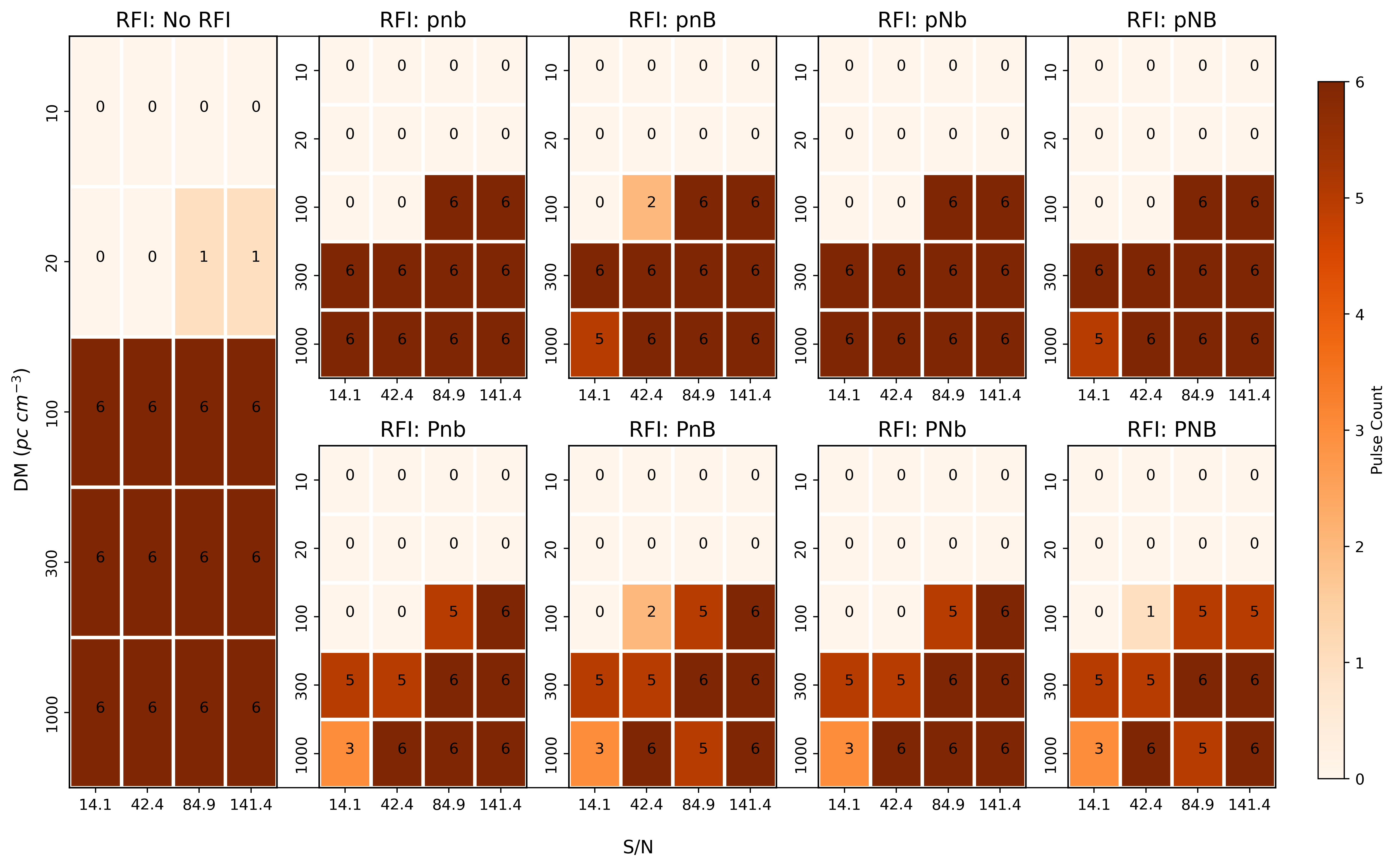}
    \caption{As for \ref{fig:realistic_iqrm_zdot_800ms_mat_plot} but for a pulse width 40 ms.}
    \label{fig:realistic_iqrm_zdot_40ms_mat_plot}
\end{figure*}

\begin{figure*}
    \centering
    \includegraphics[width=0.9\linewidth]{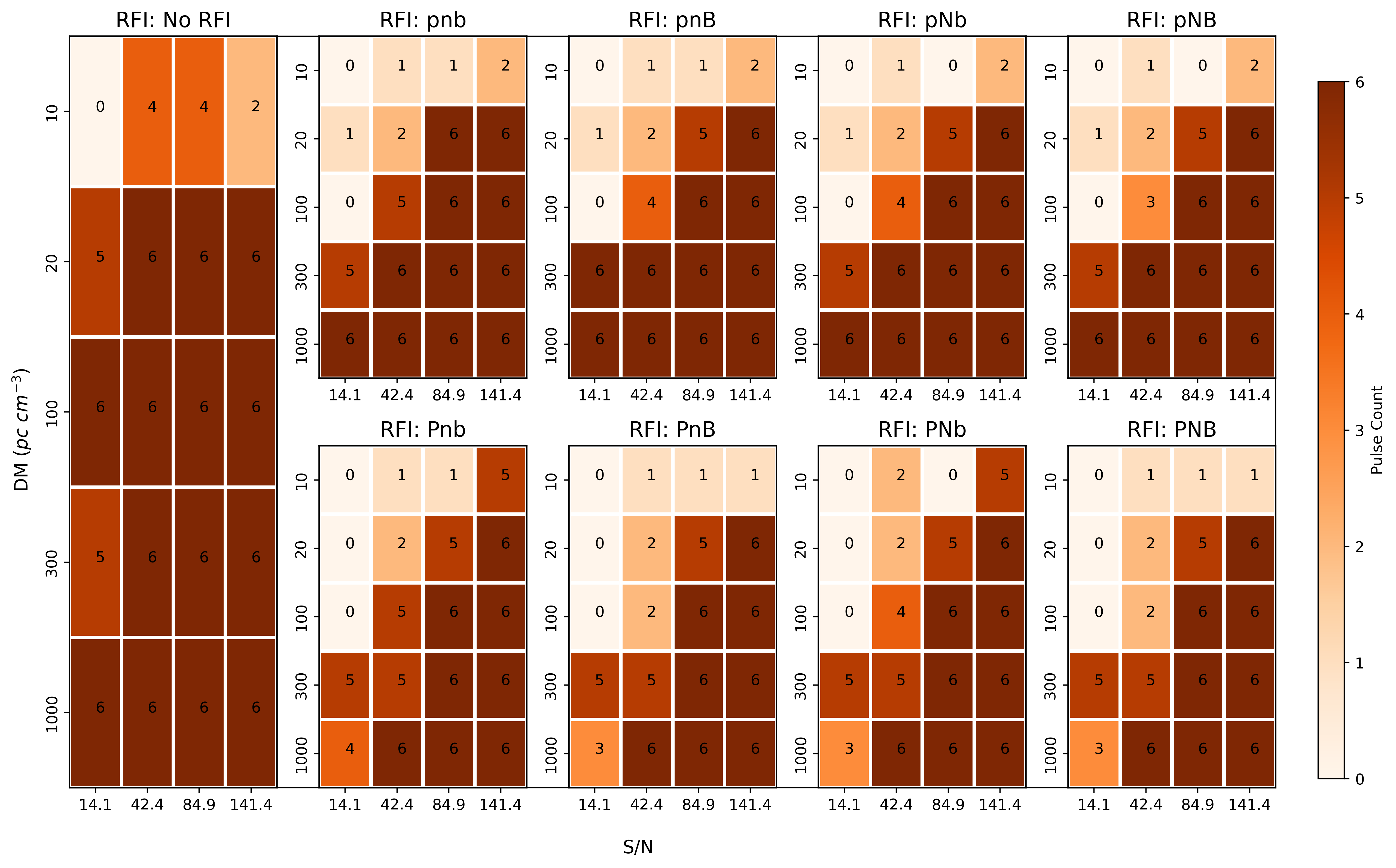}
    \caption{As for \ref{fig:realistic_iqrm_zdot_800ms_mat_plot} but for a pulse width 8 ms.}
    \label{fig:realistic_iqrm_zdot_8ms_mat_plot}
\end{figure*}

\begin{figure*}
    \centering
    \includegraphics[width=0.9\linewidth]{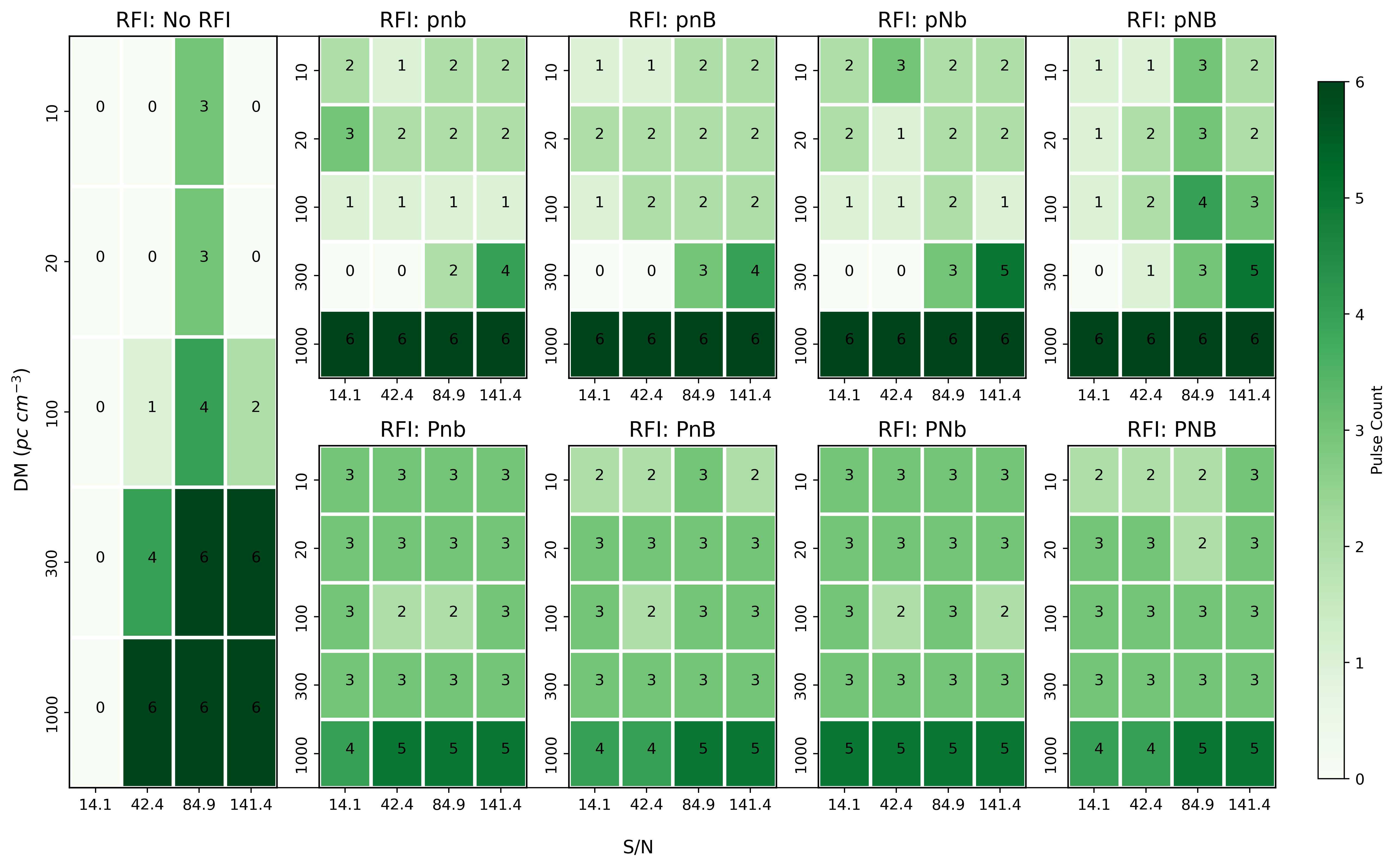}
    \caption{As for \ref{fig:realistic_iqrm_800ms_mat_plot} but SKF and ZDMF were used, and for a pulse width of 800 ms.}
    \label{fig:realistic_skf_zdot_800ms_mat_plot}
\end{figure*}

\begin{figure*}
    \centering
    \includegraphics[width=0.9\linewidth]{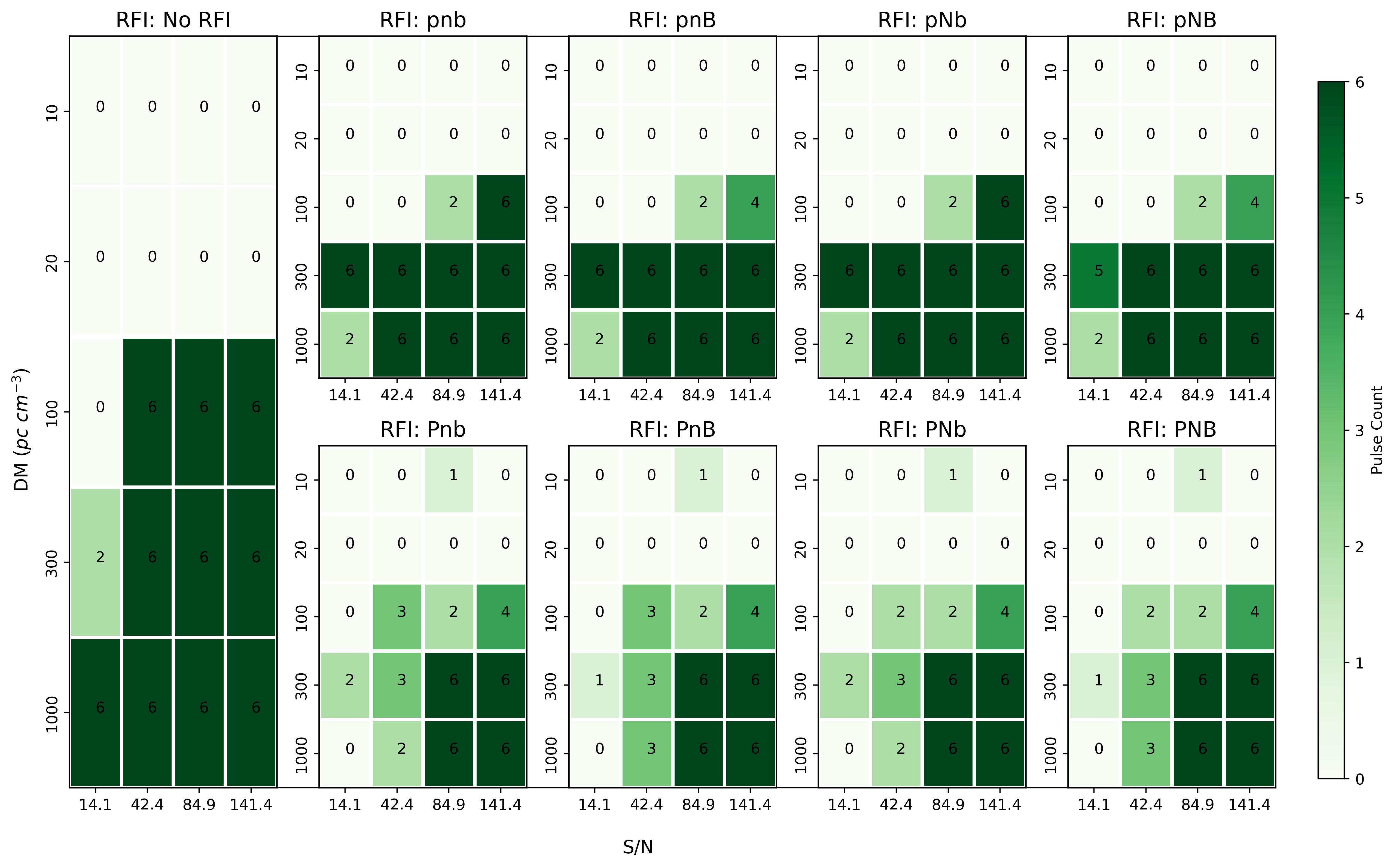}
    \caption{As for \ref{fig:realistic_skf_zdot_800ms_mat_plot} but for a pulse width 80 ms.}
    \label{fig:realistic_skf_zdot_80ms_mat_plot}
\end{figure*}

\begin{figure*}
    \centering
    \includegraphics[width=0.9\linewidth]{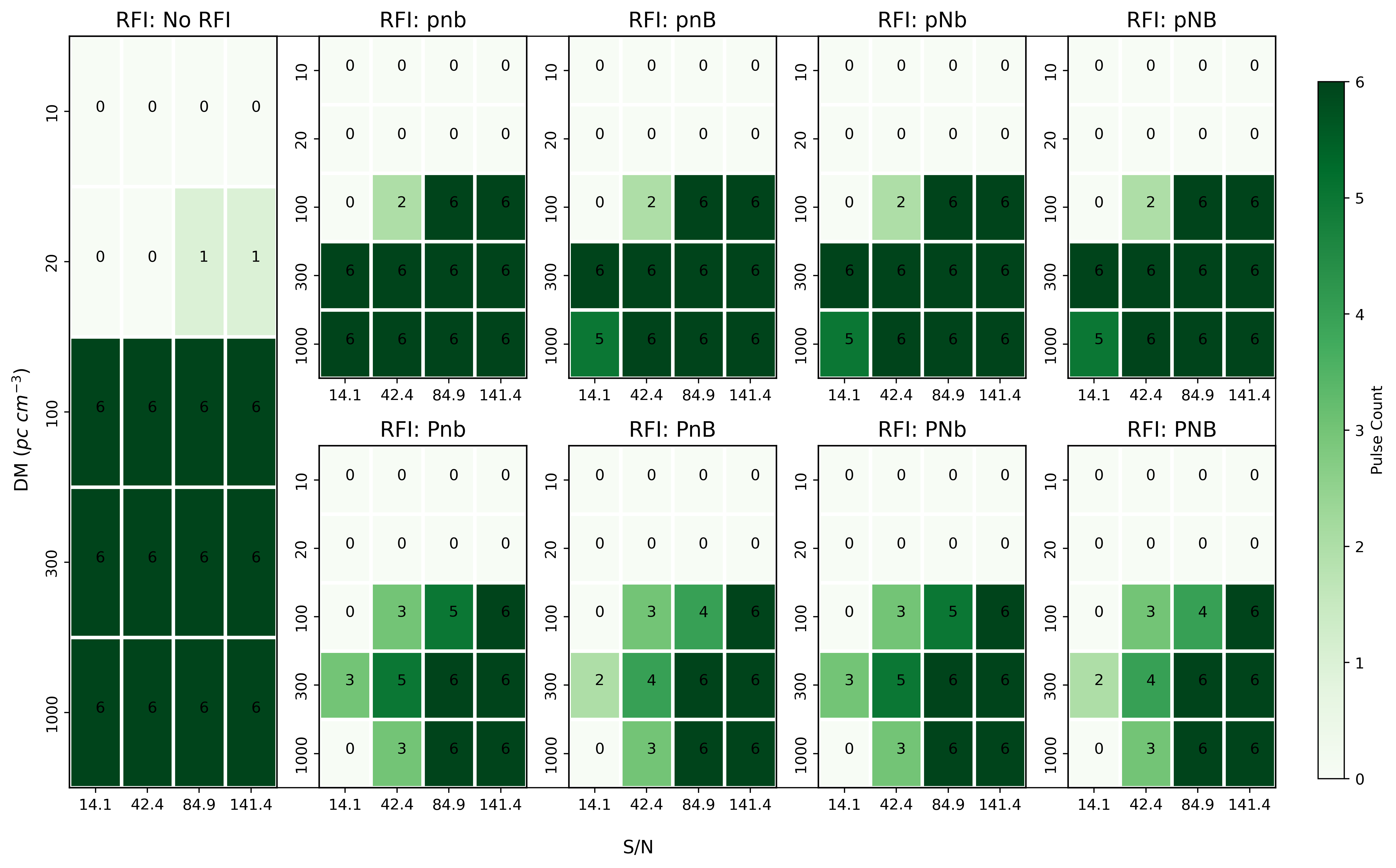}
    \caption{As for \ref{fig:realistic_skf_zdot_800ms_mat_plot} but for a pulse width 40 ms.}
    \label{fig:realistic_skf_zdot_40ms_mat_plot}
\end{figure*}

\begin{figure*}
    \centering
    \includegraphics[width=0.9\linewidth]{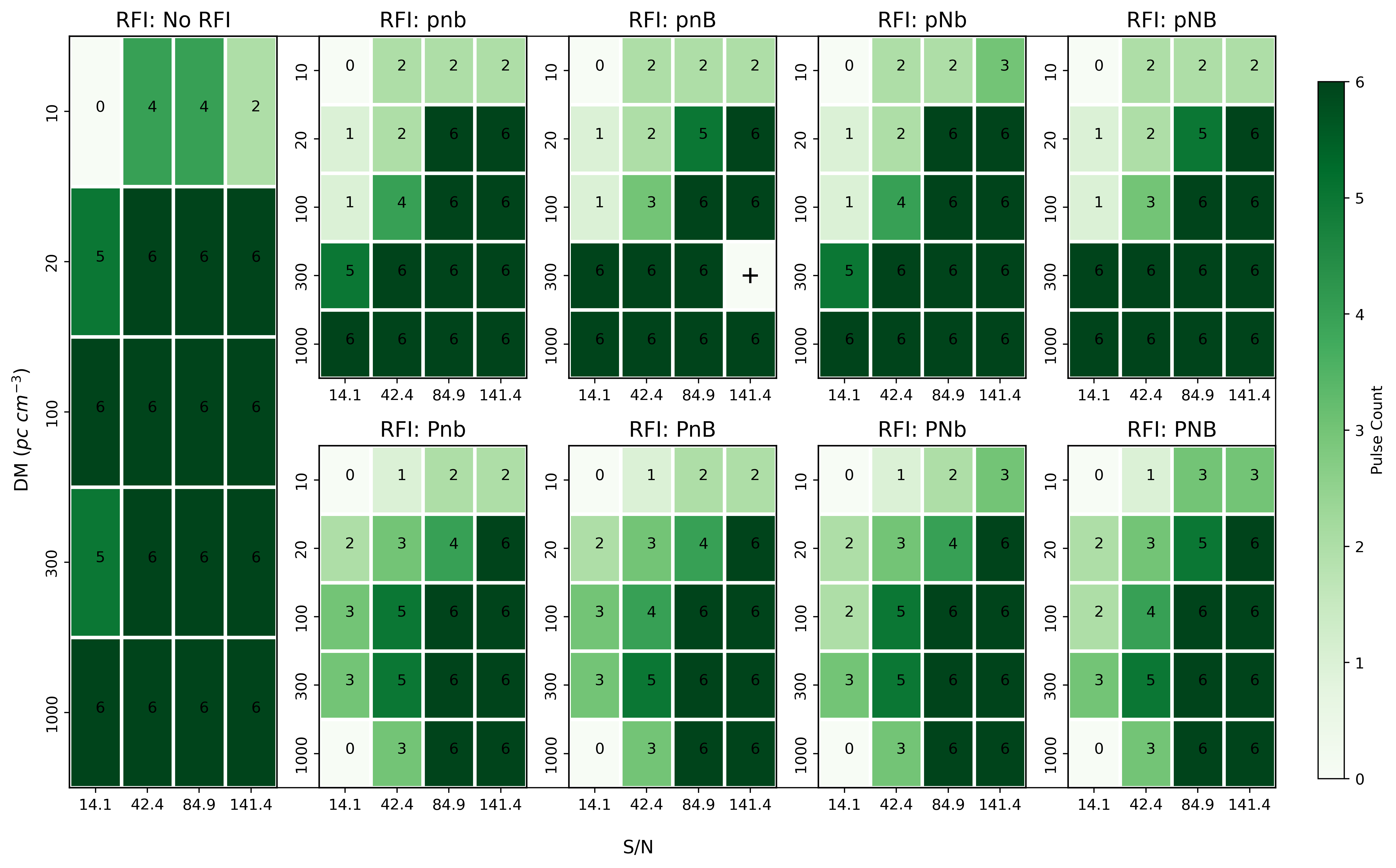}
    \caption{As for \ref{fig:realistic_skf_zdot_800ms_mat_plot} but for a pulse width 8 ms.}
    \label{fig:realistic_skf_zdot_8ms_mat_plot}
\end{figure*}

The plots below contain all the detections from all the test vectors used from TVS-2. The injected RFI is a combination of Narrowband, Periodic and Broadband RFI, with all possible cases of each type being either stronger ($\mathrm{2\sigma}$) or weak ($\mathrm{0.5\sigma}$). Fig. \ref{fig:realistic_iqrm_800ms_mat_plot}, \ref{fig:realistic_iqrm_80ms_mat_plot}, \ref{fig:realistic_iqrm_40ms_mat_plot}, \ref{fig:realistic_iqrm_8ms_mat_plot} shows the results of using IQRM on different pulse widths of the injected pulses. Fig. \ref{fig:realistic_iqrm_zdot_800ms_mat_plot}, \ref{fig:realistic_iqrm_zdot_80ms_mat_plot}, \ref{fig:realistic_iqrm_zdot_40ms_mat_plot},\ref{fig:realistic_iqrm_zdot_8ms_mat_plot} shows the results of using IQRM in sequence with ZDMF. Fig. \ref{fig:realistic_skf_zdot_800ms_mat_plot}, \ref{fig:realistic_skf_zdot_80ms_mat_plot}, \ref{fig:realistic_skf_zdot_40ms_mat_plot}, \ref{fig:realistic_skf_zdot_8ms_mat_plot} shows the results of using SKF with ZDMF. Every cell in the plots also displays the number of detections made. The marker \ding{53}~represents a case of an error by the software in exporting the candidates into a file, and $+$ represents the case of the pipeline getting timed out because of an enormous number of candidates being detected, but both \ding{53}s and $+$s, have not affected our interpretation of the results (refer to appendix \ref{sec:errors}). The latter situation is prevalent when strong broadband RFI is not fully removed and can mimic a pulse at many DMs and for many widths. The title of the subplots indicates the strength of the corresponding type of RFI in the test vectors used to generate the plot. For instance, if `PNb' is the title of the subplot then the test vectors used contain periodic and narrowband RFI of strength $2\sigma$ and broadband RFI of $0.5\sigma$ (more in Table \ref{tab:realistic_rfi_testvectors}).

\section{RFI scenario 1 and 3}
\label{severe_rfi_appendix}
\begin{figure*}
    \centering
    \includegraphics[width=\linewidth]{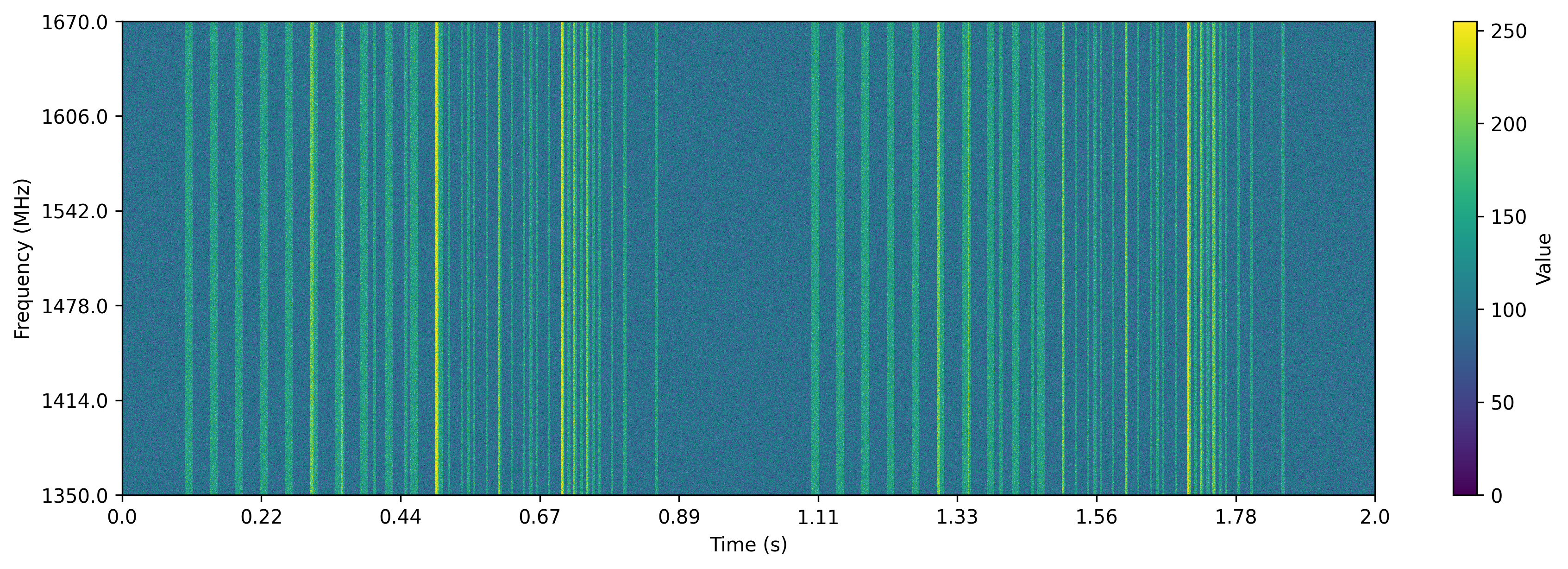}
    \caption{$1~\mathrm{s}$ long data in a test vector filterbank used, containing only periodic broadband RFI affecting 25\% of the total number of time samples in the filterbank file with a strength of $2\sigma$ (see Table~\ref{tab:extreme_rfi_testvectors}).}
    \label{fig:testvector_filterbank_broadband}
\end{figure*}

\begin{figure*}
    \centering
    \includegraphics[width=\linewidth]{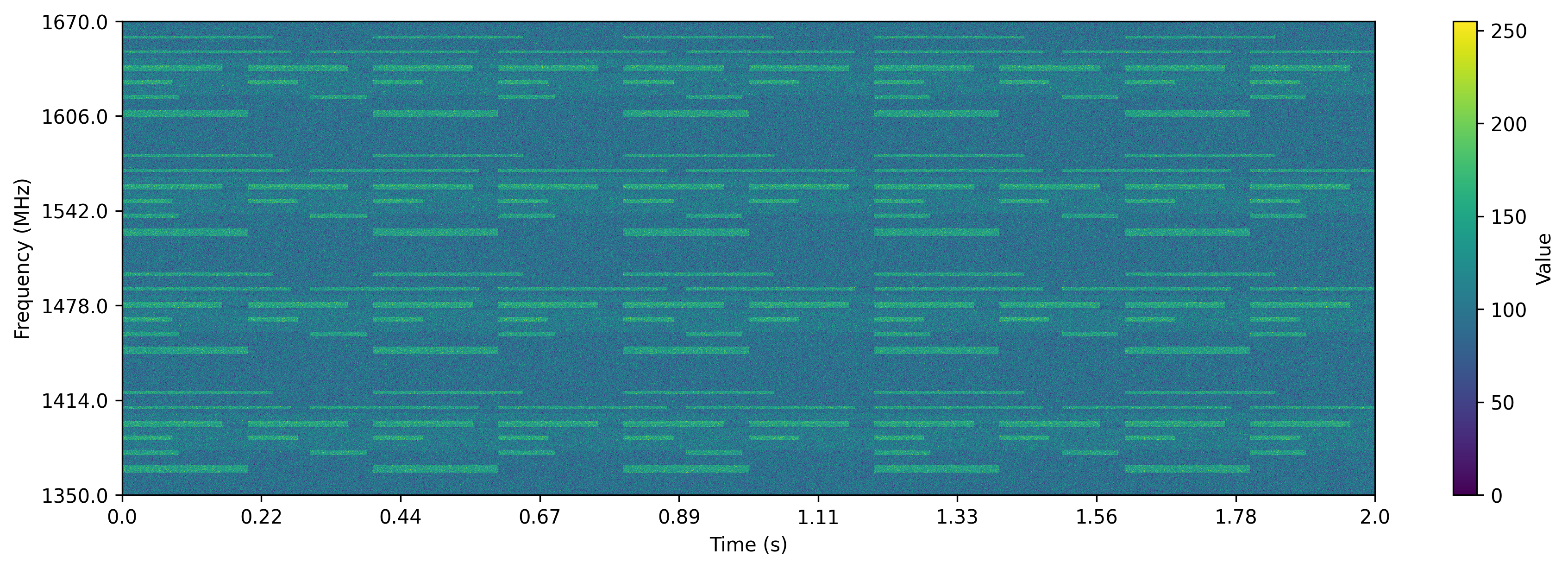}
    \caption{$1~\mathrm{s}$ long data in a test vector filterbank used, containing only narrowband RFI affecting 25\% of the total number of frequency channels in the filterbank file with a strength of $2\sigma$  (see Table~\ref{tab:extreme_rfi_testvectors}). This represents the worst-case scenario for periodic RFI.}
    \label{fig:testvector_filterbank_narrowband}
\end{figure*}

\begin{figure*}
    \centering
    \includegraphics[width=\linewidth]{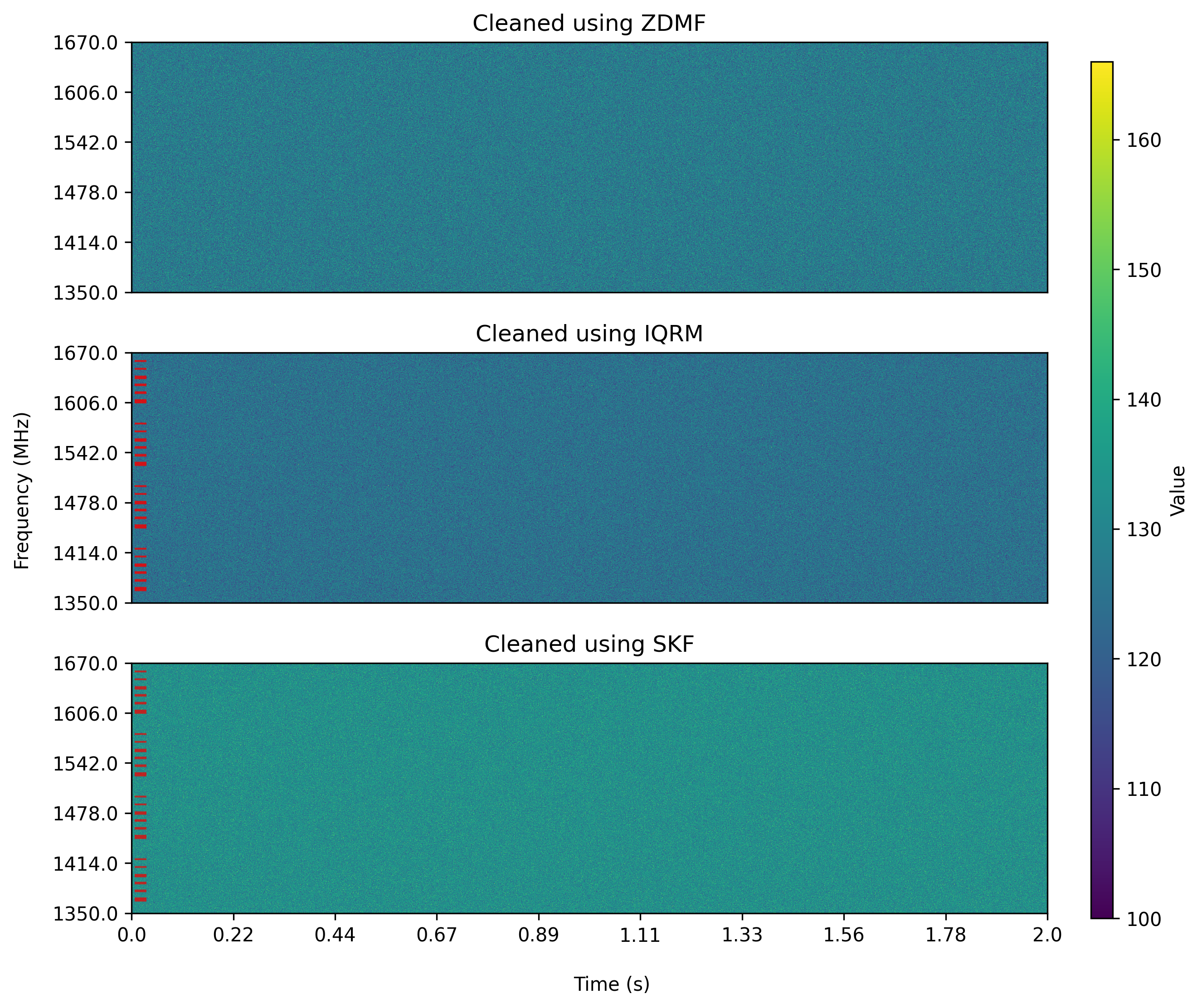}
    \caption{Demonstration of the chosen RFI removal algorithms, run individually, on the data shown in Fig.~\ref{fig:testvector_filterbank_broadband} (top panel) i.e. periodic broadband RFI of $2\mathrm{\sigma}$, cleaned using ZDMF, and Fig.~\ref{fig:testvector_filterbank_narrowband} (middle and bottom panel) i.e. periodic narrowband RFI of $2\mathrm{\sigma}$ cleaned using IQRM and SKF, similar to Fig.~\ref{fig:real_rfi_all_rfim}. The red lines at the left of the top two panels indicate the channels that are flagged as RFI-affected by the respective algorithms. filtool (used for SKF and ZDMF) corrects for the bandshape of the subset of data, whereas iqrm-apollo does not. To present the data comparable to each other, the middle panel in the figure showing data cleaned by IQRM is, therefore, also shown after correcting for its bandshape. Since ZDMF acts on the time samples, it does not mask any data but changes every time sample.}
    \label{fig:testvector_filterbank_cleaned_all_rfim}
\end{figure*}

\begin{figure*}
    \centering
    \includegraphics[width=0.9\linewidth]{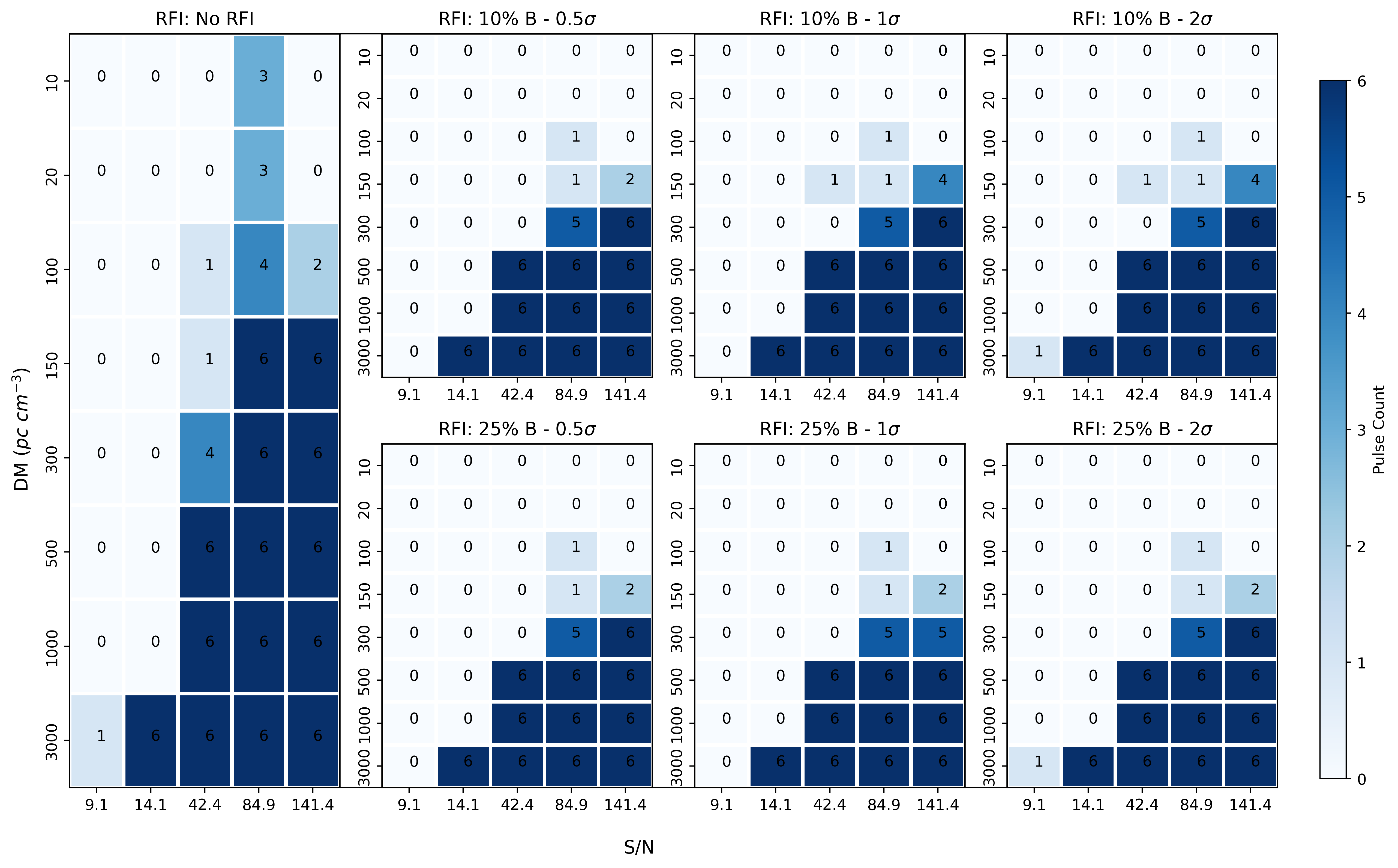}
    \caption{Number of detections of the pulses with a width of 800 ms cleaned using ZDMF as a function of DM and S/N. The left-most panel shows the no RFI cases while the other plots are for all the RFI combinations given in Table~\ref{tab:extreme_rfi_testvectors}. See Appendix \ref{sec:errors} for  the explanation of any \ding{53}'s}
    \label{fig:extreme_zdot_800ms}
\end{figure*}

\begin{figure*}
    \centering
    \includegraphics[width=0.9\linewidth]{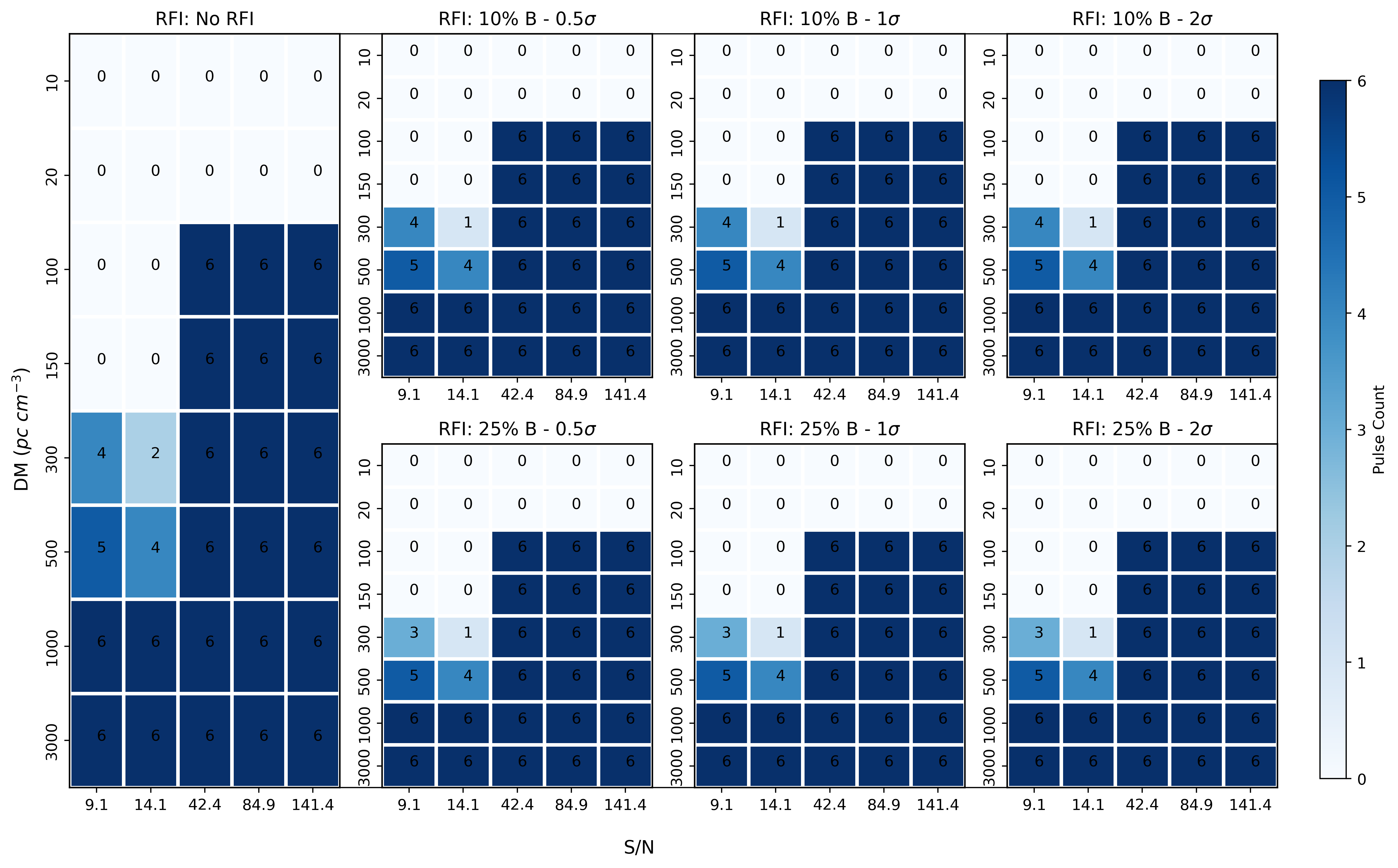}
    \caption{As for \ref{fig:extreme_zdot_800ms}, but for a pulse width 80 ms.}
    \label{fig:extreme_zdot_80ms}
\end{figure*}

\begin{figure*}
    \centering
    \includegraphics[width=0.9\linewidth]{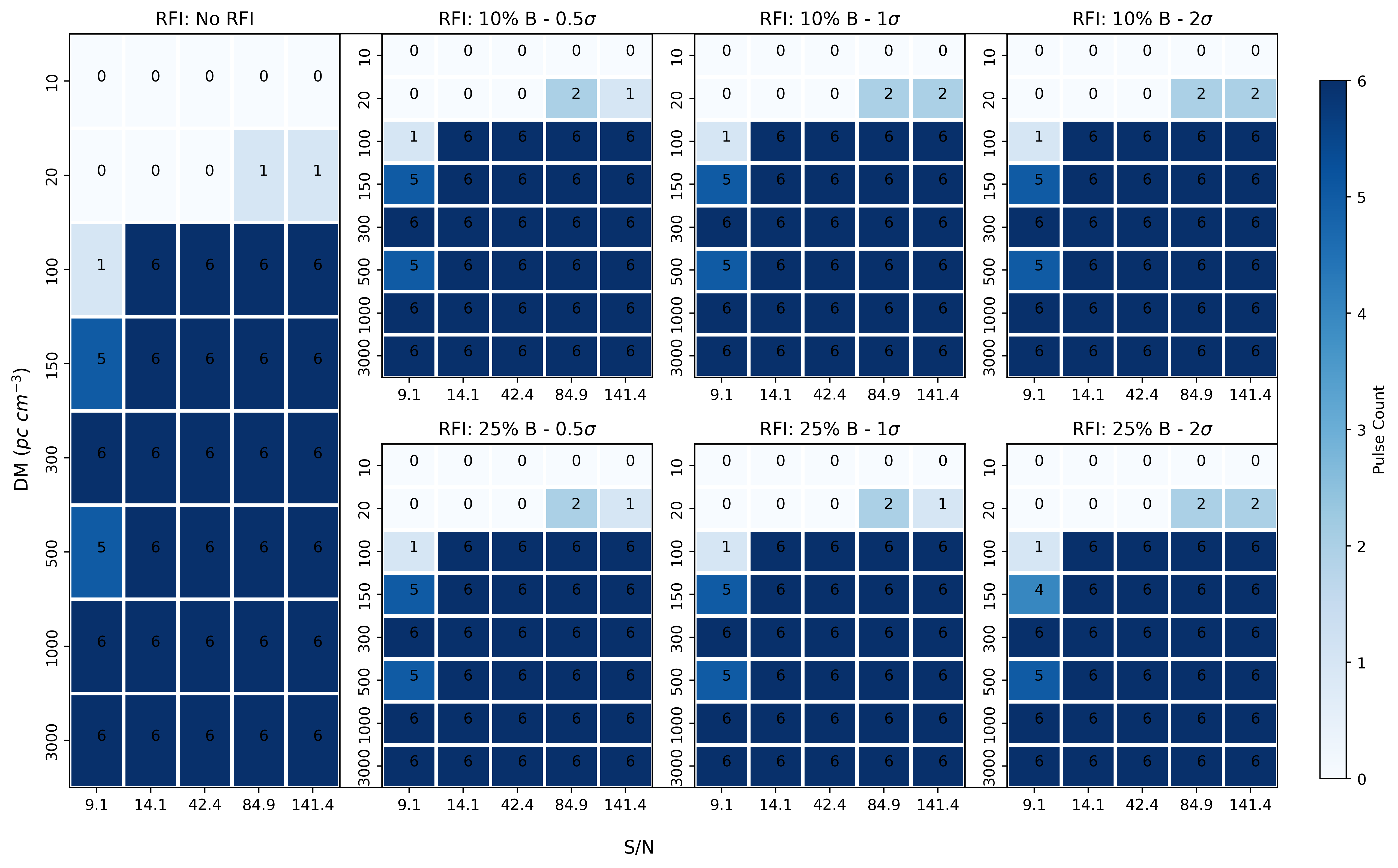}
    \caption{As for \ref{fig:extreme_zdot_800ms}, but for a pulse width 40 ms.}
    \label{fig:extreme_zdot_40ms}
\end{figure*}

\begin{figure*}
    \centering
    \includegraphics[width=0.9\linewidth]{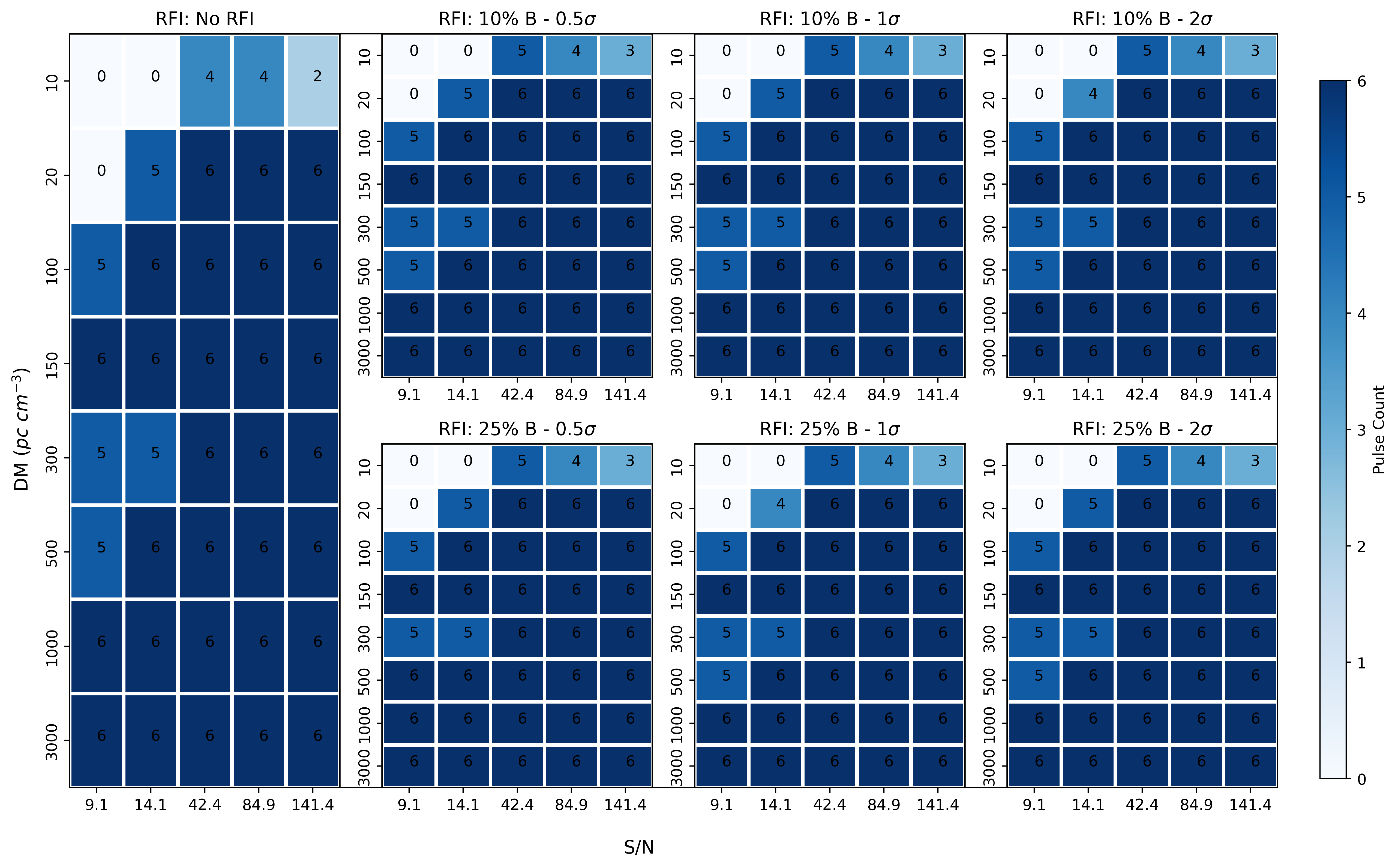}
    \caption{As for \ref{fig:extreme_zdot_800ms}, but for a pulse width 8 ms.}
    \label{fig:extreme_zdot_8ms}
\end{figure*}

\begin{figure*}
    \centering
    \includegraphics[width=0.9\linewidth]{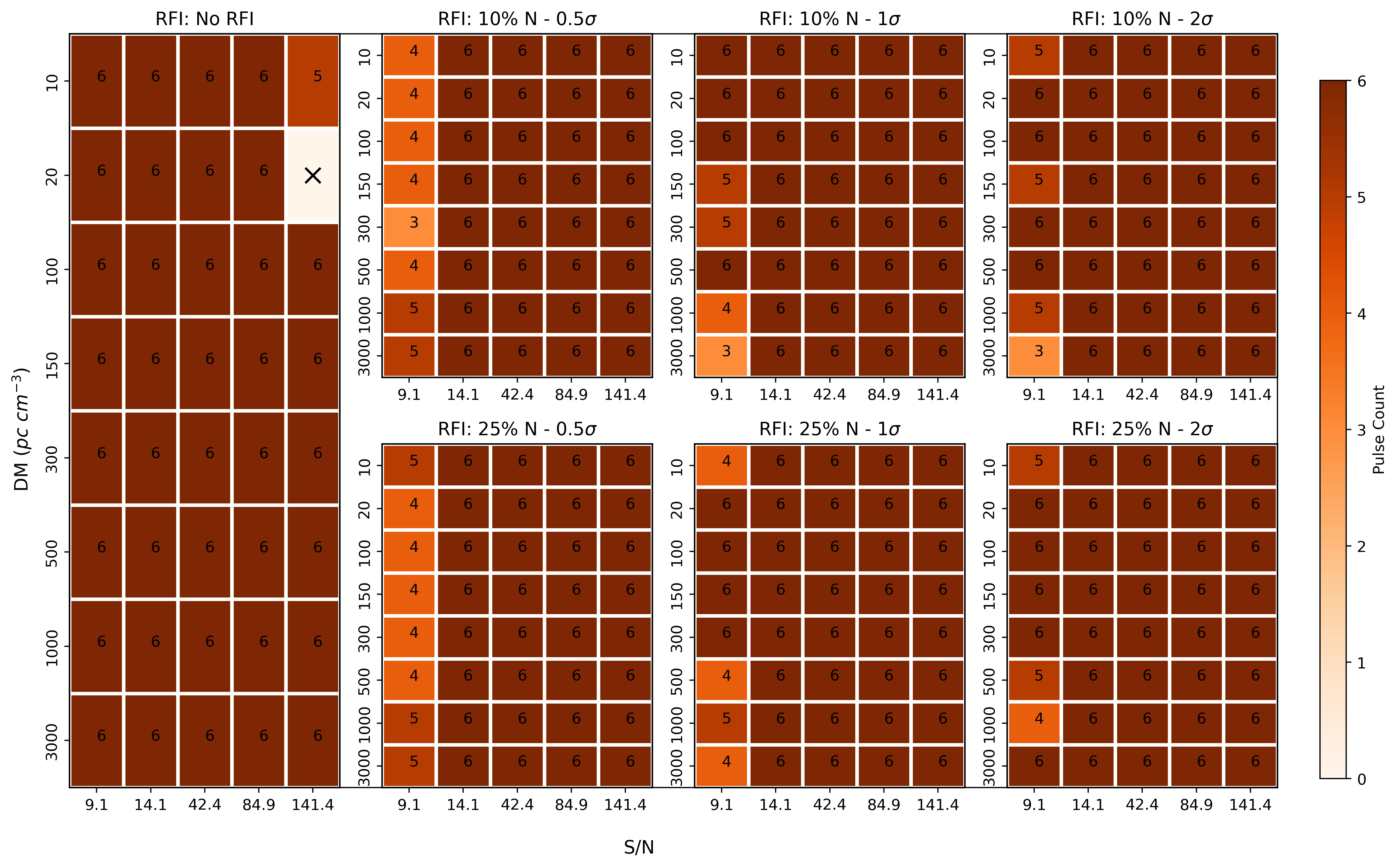}
    \caption{As for \ref{fig:extreme_zdot_800ms}, but cleaned using IQRM, and for a pulse width of 800 ms.}
    \label{fig:extreme_iqrm_800ms}
\end{figure*}

\begin{figure*}
    \centering
    \includegraphics[width=0.9\linewidth]{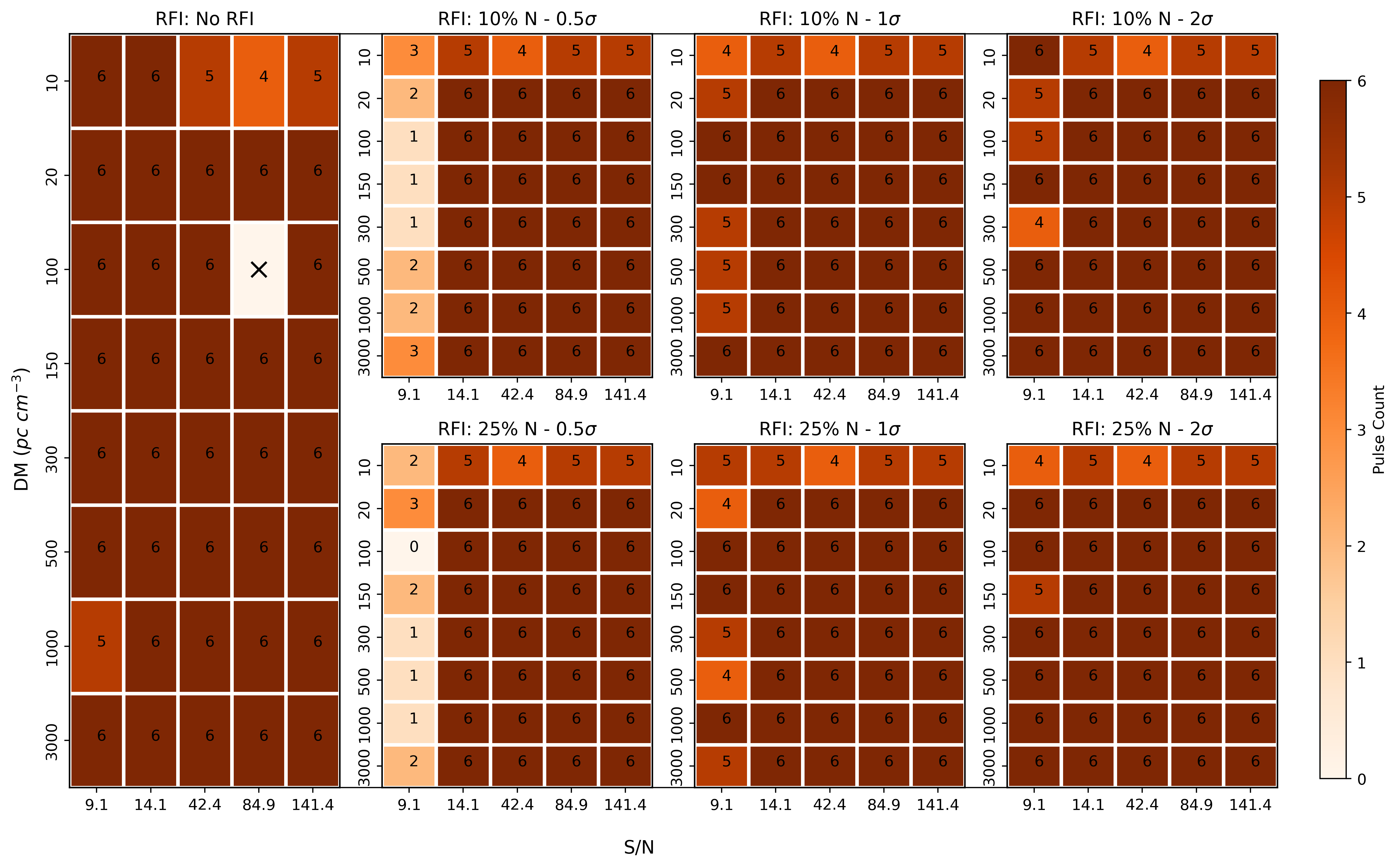}
    \caption{As for \ref{fig:extreme_iqrm_800ms}, but for a pulse width 80 ms.}
    \label{fig:extreme_iqrm_80ms}
\end{figure*}

\begin{figure*}
    \centering
    \includegraphics[width=0.9\linewidth]{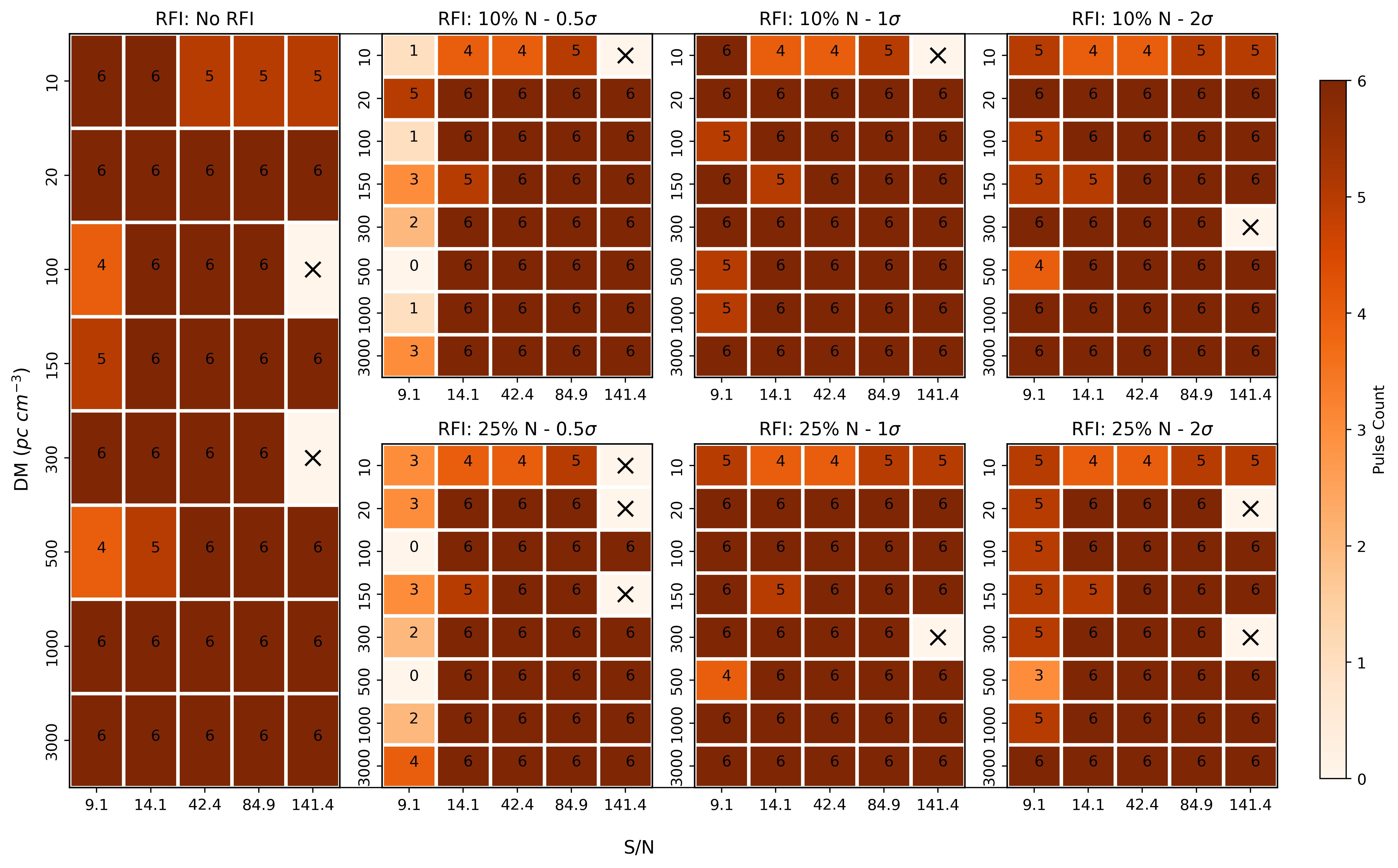}
    \caption{As for \ref{fig:extreme_iqrm_800ms}, but for a pulse width 40 ms.}
    \label{fig:extreme_iqrm_40ms}
\end{figure*}

\begin{figure*}
    \centering
    \includegraphics[width=0.9\linewidth]{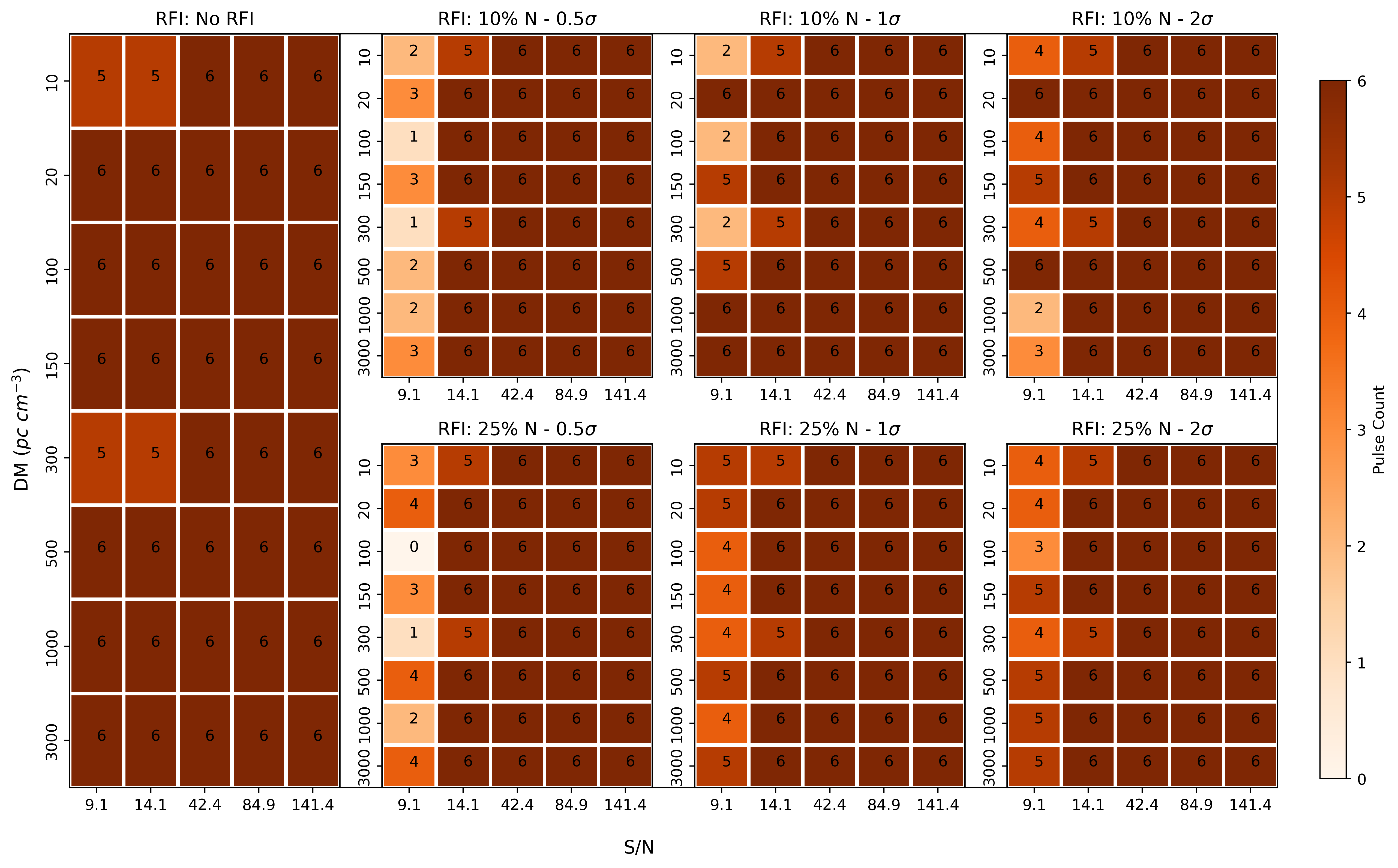}
    \caption{As for \ref{fig:extreme_iqrm_800ms}, but for a pulse width 8 ms.}
    \label{fig:extreme_iqrm_8ms}
\end{figure*}

\begin{figure*}
    \centering
    \includegraphics[width=0.9\linewidth]{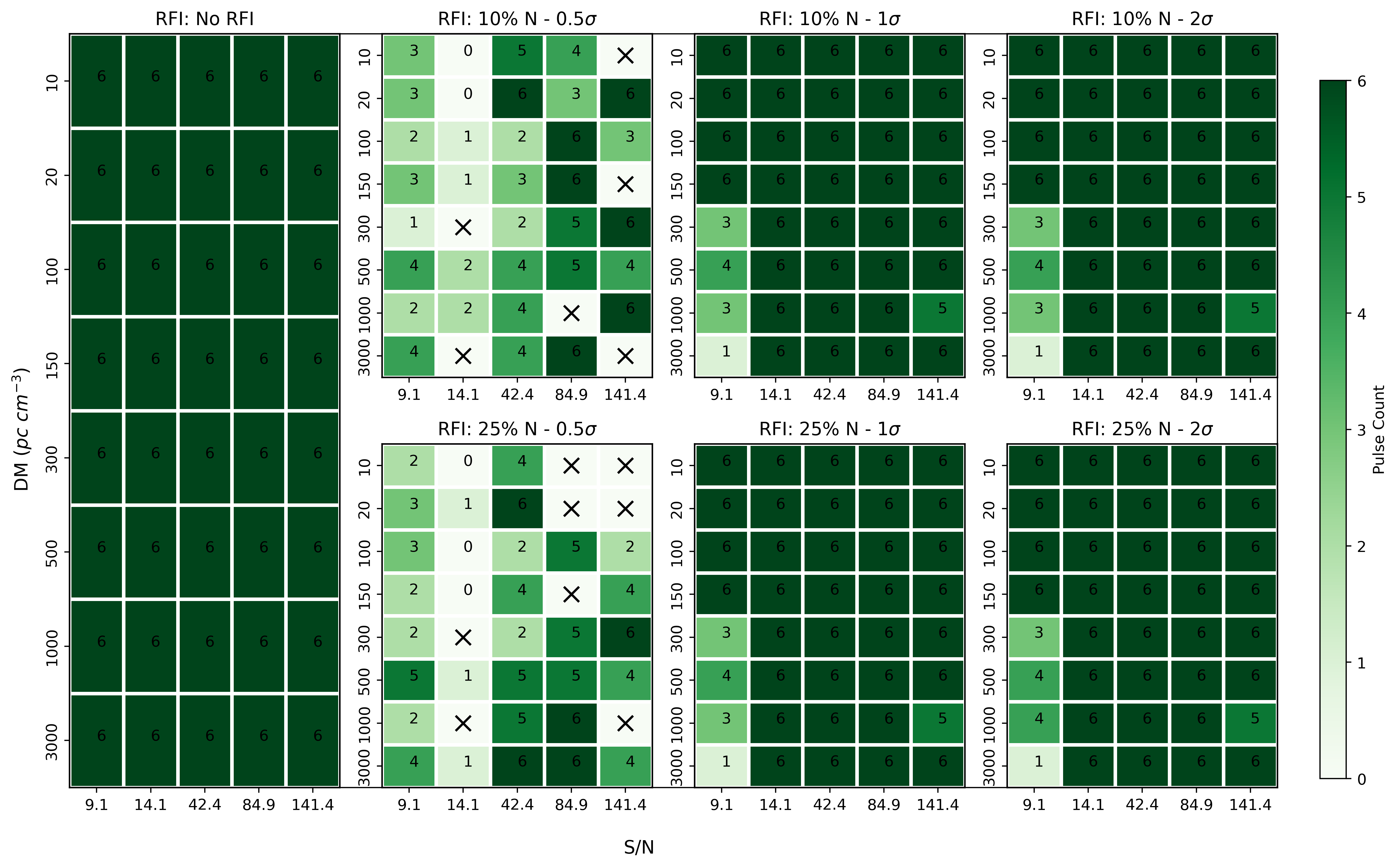}
    \caption{As for \ref{fig:extreme_zdot_800ms}, but cleaned using SKF, and for a pulse width of 800 ms.}
    \label{fig:extreme_skf_800ms}
\end{figure*}

\begin{figure*}
    \centering
    \includegraphics[width=0.9\linewidth]{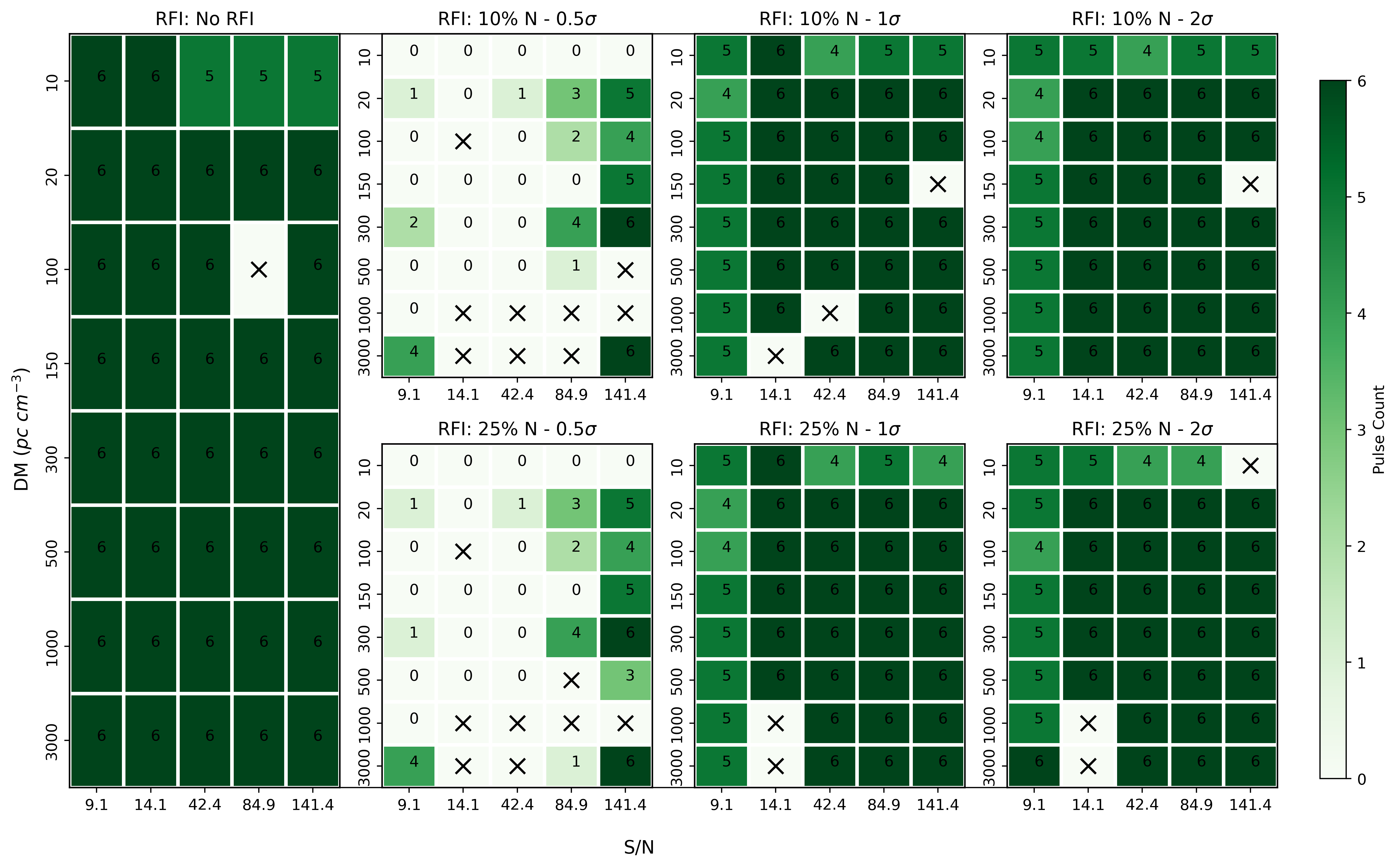}
    \caption{As for \ref{fig:extreme_skf_800ms}, but for a pulse width 80 ms.}
    \label{fig:extreme_skf_80ms}
\end{figure*}

\begin{figure*}
    \centering
    \includegraphics[width=0.9\linewidth]{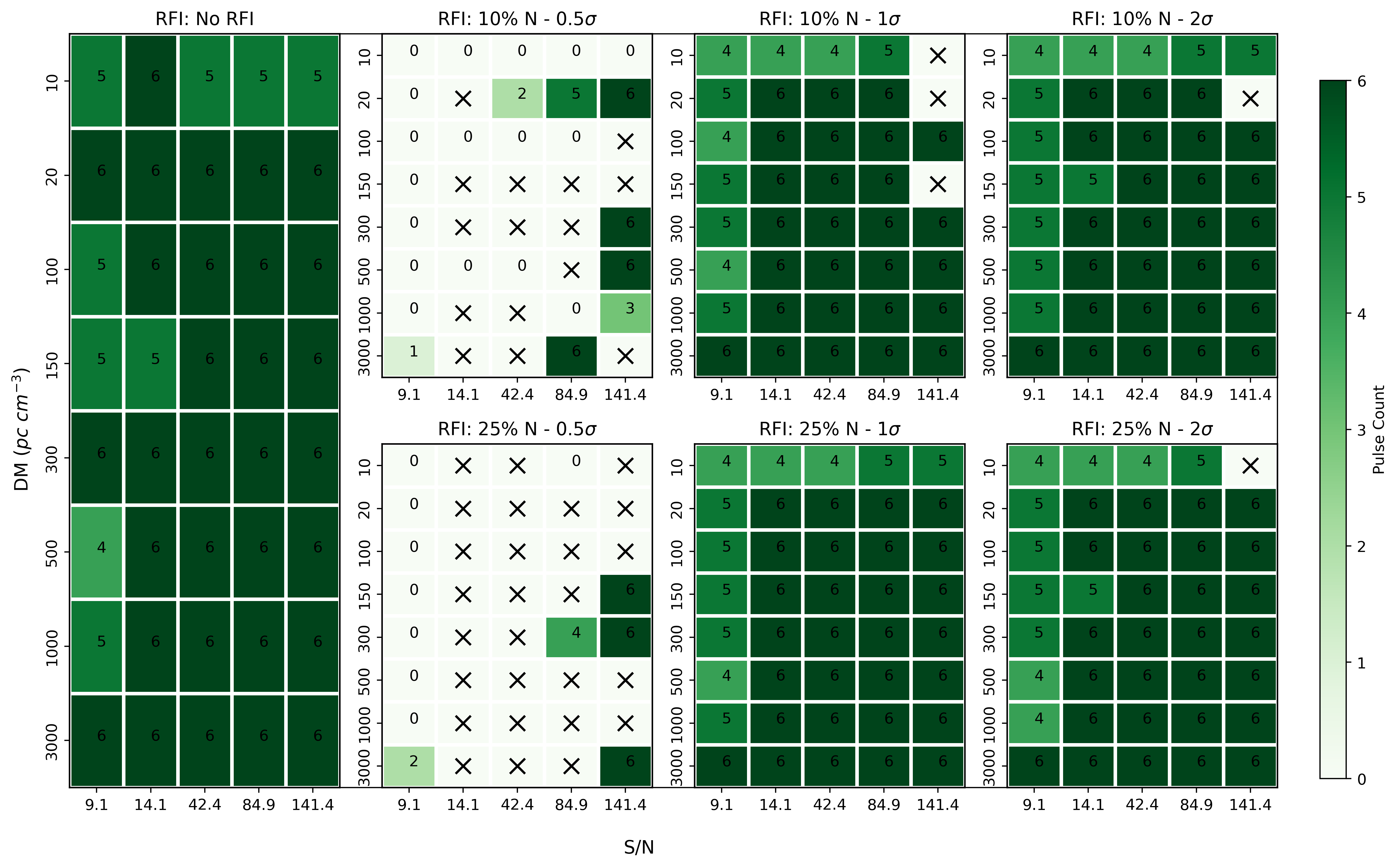}
    \caption{As for \ref{fig:extreme_skf_800ms}, but for a pulse width 40 ms.}
    \label{fig:extreme_skf_40ms}
\end{figure*}

\begin{figure*}
    \centering
    \includegraphics[width=0.9\linewidth]{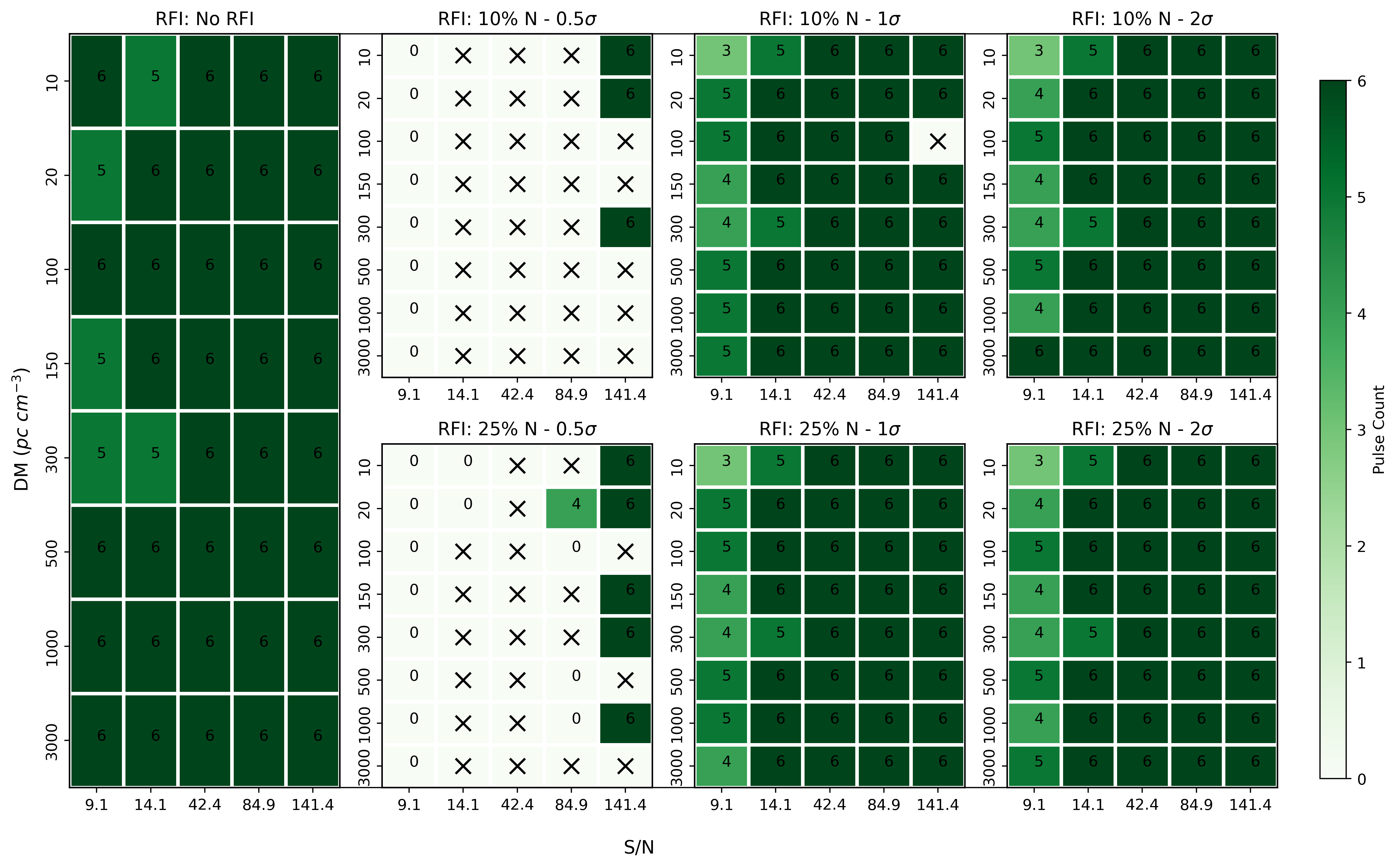}
    \caption{As for \ref{fig:extreme_skf_800ms}, but for a pulse width 8 ms.}
    \label{fig:extreme_skf_8ms}
\end{figure*}

Test vectors from TVS-1 are filterbank files containing no RFI, whereas test vectors from TVS-3 contain either only narrowband RFI or broadband RFI. Fig. \ref{fig:testvector_filterbank_broadband} and \ref{fig:testvector_filterbank_narrowband} show a chunk of data in test vectors of TVS-3 used for the tests performed. Fig.~\ref{fig:testvector_filterbank_cleaned_all_rfim} shows the same chunk of data when cleaned using ZDMF, IQRM and SKF individually. Fig.~\ref{fig:extreme_zdot_800ms}, \ref{fig:extreme_zdot_80ms}, \ref{fig:extreme_zdot_40ms}, and \ref{fig:extreme_zdot_8ms} show the number of pulses recovered when ZDMF is used to clean the test vectors containing broadband RFI. Fig. \ref{fig:extreme_skf_800ms}, \ref{fig:extreme_skf_80ms}, \ref{fig:extreme_skf_40ms}, and \ref{fig:extreme_skf_8ms} show the results of the number of detections when SKF is used to clean the test vectors containing narrowband RFI. Fig. \ref{fig:extreme_iqrm_800ms}, \ref{fig:extreme_iqrm_80ms}, \ref{fig:extreme_iqrm_40ms}, and \ref{fig:extreme_iqrm_8ms} show the results of the number of detections when IQRM is used to clean the test vectors containing narrowband RFI. The \ding{53}~ here represents the same as defined in the Appendix \ref{realistic_rfi_appendix}.

\bsp	
\label{lastpage}
\end{document}